\documentclass[12pt]{iopart}

\expandafter\let\csname equation*\endcsname\relax
\expandafter\let\csname endequation*\endcsname\relax
\usepackage{amsmath}

\usepackage{iopams}  
\usepackage{graphicx}
\usepackage{siunitx}
\usepackage{booktabs}
\usepackage{xcolor}
\usepackage{cite}  
\usepackage[hidelinks]{hyperref}

\usepackage{breqn}

\usepackage[explicit]{titlesec}
\usepackage{ulem}
\usepackage{lipsum}% just to generate text for the example
\usepackage{array}

\DeclareMathOperator{\JSD}{JSD}

\newcolumntype{P}[1]{>{\centering\arraybackslash}p{#1}}

\titleformat{\paragraph}[runin]
  {\normalfont\normalsize\bfseries}{}{15pt}{{\theparagraph\hspace*{1em}#1.}}
\titleformat{name=\paragraph,numberless}[runin]
  {\normalfont\normalsize\bfseries}{}{15pt}{{#1.}}

\begin{document}

\title{New Angles on Fast Calorimeter Shower Simulation }

\vspace{10pt}

\author{Sascha Diefenbacher$^1$, Engin Eren$^2$, Frank Gaede$^{2,3}$, Gregor Kasieczka$^{1,3}$, Anatolii Korol$^2$, Katja Kr\"uger$^2$, Peter McKeown$^{2*}$, Lennart Rustige$^{2,3}$}

\vspace{10pt}

\address{$^1$ Institut f\"ur Experimentalphysik, Universit\"at Hamburg, Luruper Chaussee 149, 22761 Hamburg, Germany} 
\address{$^2$ Deutsches Elektronen-Synchrotron DESY, Notkestr. 85, 22607 Hamburg, Germany} 
\address{$^3$ Center for Data and Computing in Natural Sciences CDCS, Deutsches Elektronen-Synchrotron DESY, Notkestr. 85, 22607 Hamburg, Germany}
\vspace{10pt}
\address{$^*$ Author to whom correspondence should be addressed.}
 \vspace{-1pt}  
\ead{peter.mckeown@desy.de}
\vspace{10pt}
\begin{indented}
\item[]March 2023
\end{indented}

\begin{abstract}
The demands placed on computational resources by the simulation requirements of high energy physics experiments motivate the development of novel simulation tools. Machine learning based generative models offer a solution that is both fast and accurate. In this work we extend the Bounded Information Bottleneck Autoencoder (BIB-AE) architecture, designed for the simulation of particle showers in highly granular calorimeters, in two key directions. First, we generalise the model to a multi-parameter conditioning scenario, while retaining a high degree of physics fidelity. In a second step, we perform a detailed study of the effect of applying a state-of-the-art particle flow-based reconstruction procedure to the generated showers. We demonstrate that the performance of the model remains high after reconstruction. These results are an important step towards creating a more general simulation tool, where maintaining physics performance after reconstruction is the ultimate target.
\end{abstract}

\newpage

\section{Introduction}
The detailed simulation of particle interactions with sophisticated detector systems is central to modern high energy physics, providing both the bridge between theory and experiment and the means to design and optimise detectors for future experiments. Traditionally, simulation tools have been based on the use of Monte-Carlo methods, with \textsc{Geant4}~\cite{g4} being the most prevalent. This approach can produce very high quality simulations, but comes at the price of consuming significant computational resources~\cite{HEPSoftwareFoundation:2017ggl}. 

A promising alternative approach to potentially speed up simulation is to use a generative model based surrogate simulator. To this end, a plethora of different generative models have been proposed for the task, including Generative Adversarial Networks (GANs)~\cite{CaloGAN, CaloGAN3, Khattak:2021ndw, AtlFast3, Rehm:2021zoz, kansal2021graph, EPIC, IEA_GAN}, Bounded Information Bottleneck Autoencoders (BIB-AEs)~\cite{GettingHigh}~\cite{HBFS}, Wasserstein GANs (WGANS)~\cite{ErdmannWGAN}~\cite{ErdmannWGAN2}, Normalising Flows~\cite{Krause:2021Flow1}~\cite{Krause:2021FLow2}~\cite{L2L} and Score-Based Models~\cite{Mikuni:2022Score}.

A major challenge for generative model based fast simulation approaches is that the next generation of particle physics experiments that will be performed at colliders feature detectors of an ever increasing granularity. This is for example true of both the Calorimeter Endcap Upgrade for the CMS experiment in preparation for the upcoming High Luminosity phase of the LHC~\cite{CMS_HGCAL}, and for experiments at so called Higgs Factories. Such a facility would consist of an $e^{+}e^{-}$ collider, being either circular or linear, that would provide clean environments to enable high precision studies of the Higgs, electroweak and top sectors~\cite{ILC_Snowmass}~\cite{FCC_Snowmass}. The high granularity present in these detectors places high demands on the physics performance of a generative model based simulator, as well as presenting data of a high dimensional nature.

Building on previous studies into the simulation of both electromagnetic~\cite{GettingHigh}~\cite{decoding_photons} and hadronic~\cite{HBFS} showers in a highly granular calorimeter with a highly performant BIB-AE model, the contribution of this work is twofold. First, we extend the previous studies by demonstrating the ability to condition a BIB-AE on multiple parameters, while maintaining a high degree of physical fidelity with respect to \textsc{Geant4}. This is a crucial step towards being able to apply such simulators in a realistic environment, where particles can impact a detector with various angles as well as energies. Secondly, we perform a detailed study of the effect of applying a state-of-the-art reconstruction algorithm and demonstrate that the BIB-AE based simulator maintains a strong performance. This evaluation serves as a strong indicator of the quality of a model, which will ultimately be judged on its physics performance after reconstruction.

The paper is organised as follows. Section \ref{sec:Dat_Rec} gives details on the dataset and describes the particle flow reconstruction algorithm used. The generative model\footnote{The implementation of the model and the hyperparameters used for training are available at \url{https://github.com/FLC-QU-hep/new_angles}} is described in Section~\ref{sec:models}, and a review of its performance before and after reconstruction, as well its computational performance, is shown in Section~\ref{sec:Results}.

\section{Dataset and Reconstruction Scheme}\label{sec:Dat_Rec}
In this section, the calorimeter system used in this study, and the production of the dataset, is described in Section \ref{sec:Dataset}, while the particle flow algorithm used for reconstruction is outlined in Section \ref{sec:Reco}. 

\subsection{Dataset} \label{sec:Dataset}

For this work, we focus on the International Large Detector (ILD)~\cite{ILD-IDR}, a detector concept proposed for the the International Linear Collider (ILC), which is one option for a Higgs Factory and a high energy lepton collider. The ILD detector is optimized for the particle flow algorithm, which aims to reconstruct each individual particle in an event, with the goal of obtaining the best overall detector resolution possible. 

This introduces a number of requirements on the detector --- the most relevant to this study is the high granularity of the calorimeters required. The subject of this work is the simulation of photon showers in the Si-W option for the ILD electromagnetic calorimeter (ECAL). This detector is a sampling calorimeter which consists of 30 layers of active silicon sensors sandwiched between tungsten absorber layers. It features two sampling fractions --- with the first $20$ tungsten layers being $2.1$ mm thick, while the last $10$ layers have a thickness of $4.2$ mm. The silicon layers feature cells of size $5\times5$ mm$^2$.

The iLCSoft~\cite{ilcsoft} ecosystem is used by ILD for simulation and reconstruction of the detector response, as well as for subsequent analysis. The training data used in this work is produced via full simulation with \textsc{Geant4} %~\cite{g4} 
version $10.4$ using the \textrm{QGSP\_\,BERT} physics list, using a realistically detailed detector model implemented in \textsc{DD4hep}~\cite{dd4hep}. The training dataset consists of showers in the ILD ECAL, which are initiated by photons fired from a virtual \textit{particle gun}. This \textit{particle gun} has a fixed position of $(x', y' z') = (0.0 \; \mathrm{mm}, 1810 \; \mathrm{mm}, -50 \; \mathrm{mm})$\footnote{The gun position is chosen such that the particles are created right at the face of the ECAL and do not hit gaps in the calorimeter} in the ILD coordinate system. This coordinate system is orientated such that $z'$ points along the beam axis, and $y'$ points vertically, i.e. perpendicularly to the calorimeter face. The shower images used for training are produced by projecting the ECAL hits into a regular grid of $30\times60\times30$ voxels in the local $(x,y,z)$ coordinate system. In these transformed coordinates, the $z$ axis lies orthogonal to the face of the calorimeter, so each plane of voxels along this dimension corresponds to one of $30$ layers in the physical ECAL. In this projected space, the photons are incident at a fixed cell at index $(i_x, i_y, i_z) = (15, 12, 0)$. Since the geometry of the ECAL is not perfectly regular, a staggered pattern tends to appear along the $z$ direction of the images~\cite{GettingHigh}. The projection from the physical geometry to a regular grid also tends to cause artefacts which are corrected for such that each voxel in the regular grid corresponds to exactly one sensor in the calorimeter. A total of $500$k showers are present in the training dataset, produced by photons with varying incident energies and angles\footnote{The training dataset is available at \url{https://doi.org/10.5281/zenodo.7786846}}. The incident angle is varied in one direction such that it corresponds to the polar angle in the global ILD coordinate system (i.e. the angle with respect to the $z'$-axis). The incident energy varies uniformly in the range of $10$ to $100$ GeV, along with the angle in the $y-z$ plane, which simultaneously varies from $90$ to $30$ degrees to the calorimeter face. The angle in the local $x-z$ plane (corresponding to the azimuthal angle in the global ILD coordinate system) is fixed to be $90$ degrees. Finally, $9$ test datasets, each consisting of $1900$ showers produced by photons at fixed combinations of energies (\{20, 50, 90\}~GeV)  and polar angles (\{40, 60, 85\} degrees), are used to check the model performance across the phase space. This allows the evaluation of the single energy and angular response of the model, which is the end target of a simulator.
%For this purpose, combinations of incident energies of \{20, 50, 90\} GeV with incident polar angles of \{40, 60, 85\} degrees were made.
%\PJM{TO DO: Should we give some reasoning why we only varied the polar angle in one direction here?- something like we wanted to extend the model to handle multi-parameter conditioning, and to get a wide range of angular coverage, without increasing the size of the 3D calorimeter grid too drastically} \fg{I think this is not needed here...}
A fixed calibration factor is applied to scale the hit energies in the last 10 layers to account for the change in the sampling fraction arising from the thicker absorber layers.

\subsection{Reconstruction Scheme} \label{sec:Reco}
The precision physics goals of future $e^{+}$$e^{-}$ Higgs factories require the ubiquitous adoption of particle flow reconstruction at these facilities. This approach aims to reconstruct each individual particle in an event using information from the best subdetector system for the given task, in order to optimise the overall detector resolution. In this work, we make use of $\textsc{PandoraPFA}$~\cite{pandora_PFA}, the state-of-the-art pattern recognition algorithm used by ILD and other linear collider detector concepts.

The full reconstruction chain used by ILD is a multi-step process. Firstly, a digitisation procedure is applied to the hits produced by the simulation. This involves emulating effects which are for instance intrinsic to the sensor or arise from the readout electronics. Secondly, for the calorimeters a two step calibration procedure is applied to convert the total visible energy deposited in the detector back to the incident particle's energy in GeV.
Next, after several pattern recognition algorithms to reconstruct the tracks from charged particles are run, sophisticated clustering procedures are iteratively applied to the resulting calorimeter hits by PandoraPFA, using track information where appropriate. The final result is a list of reconstructed particles, or Particle Flow Objects (PFOs), with important information pertaining to the particles (energy, momentum, ID, etc.) associated. This associated information is then used for high level reconstruction and subsequent physics analysis.

\section{Generative Models} \label{sec:models}

\subsection{Bounded Information Bottleneck Autoencoder} \label{sec:BIBAE}
%The Information Bottleneck (IB) Method \cite{IB_Method} was formulated to provide a theoretical principle by which the information from a given random variable $X \in \mathcal{X}$, relevant for the prediction of another random variable $Y \in \mathcal{Y}$, could be extracted. It has since played an important role in approaches seeking to gain an information-theoretical understanding of Deep Learning ~\cite{IB_DNN}.

%In the framework of the IB a purely unsupervised ($U$) task in which no labels are provided can be considered to be a compression from the data space $X$ to a lower dimensional latent space $Z$, which seeks to maximise the mutual information $I$ between $Z$ and $X$, while minimizing information irrelevant for the task. The problem can be formulated as a minimization of the Lagrangian

%\begin{equation}
%    \mathcal{L}^{U}(\phi)= I_{\phi}(X;Z) - \beta I(Z;X)
%\end{equation}
%over model parameters $\phi$, in which the Lagrange multiplier $\beta$ parameterizes the trade off between compression and retention of useful information.  

\begin{figure}[t]
    \centering
    \includegraphics[width=0.9\textwidth]{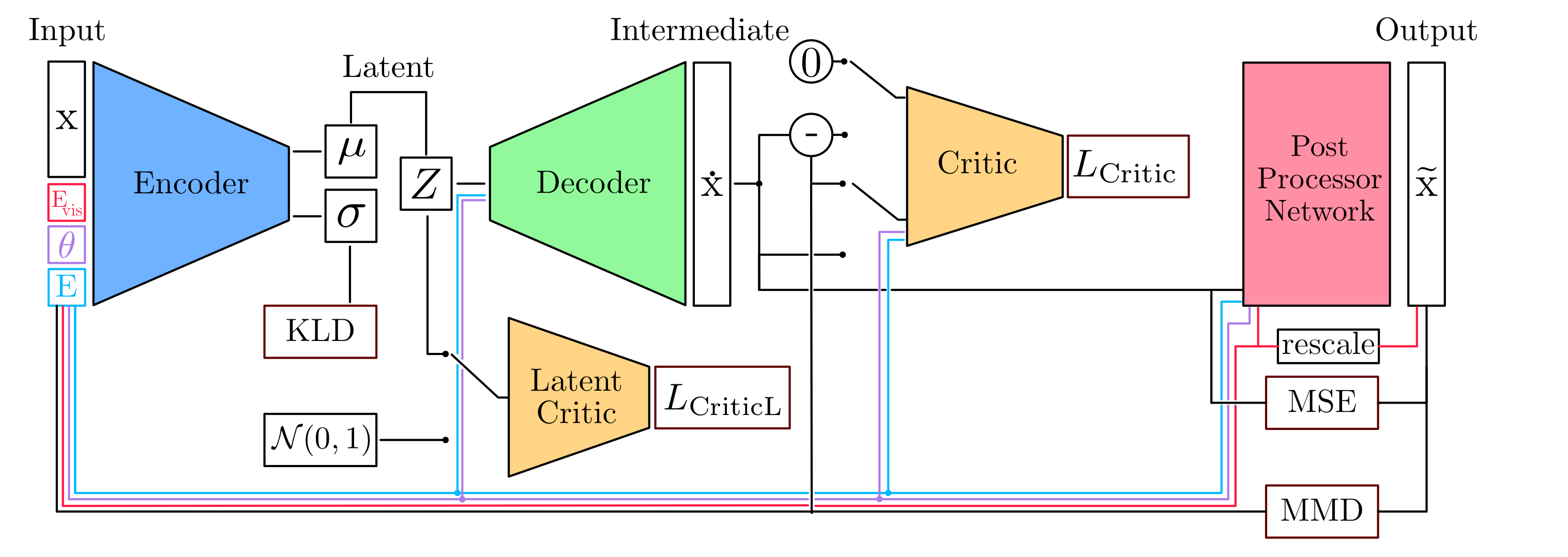}    %{BIBAE_3_Angle.pdf}
    \caption{Schematic diagram of the BIB-AE architecture setup during training, including each network and its corresponding loss terms. The encoder reduces the input calorimeter showers to a low-dimensional latent space, which is regularised by KLD and MMD loss terms in addition to a latent critic. The decoder reconstructs shower images back from the latent space, with a dual purpose reconstruction critic simultaneously assisting reconstruction and judging the quality of the output shower. The Post Processor network is trained in a second step to adjust voxels individually. The blue and lilac lines represent an input conditioning on the energy and angle of the incident particle, while the red line represents a conditioning on the visible energy.}
    \label{fig:BIBAE_PP_arch}
\end{figure}

%The Bounded Information Bottleneck Autoencoder (BIB-AE) first introduced in ~\cite{BIB-AE}, resulted from an analysis of unsupervised generative models from the perspective of the IB. 
The Bounded Information Bottleneck Autoencoder (BIB-AE), first introduced in~\cite{BIB-AE}, generalises many of the features present in common Generative Adversarial (GAN) and Variational Autoencoder (VAE) architectures. The BIB-AE architecture was successfully applied to the problem of calorimeter shower simulation in~\cite{GettingHigh}, with further improvements in~\cite{decoding_photons} and~\cite{HBFS}. The \textsc{PyTorch}~\cite{pytorch} implementation described here extends the one developed in~\cite{HBFS}, and is illustrated in Figure \ref{fig:BIBAE_PP_arch}.

At its core, the BIB-AE is an autoencoder, with an encoder network $N_E$ mapping calorimeter showers to a lower dimensional latent space, and a decoder network $N_D$ reconstructing showers from the latent representation. Around this core, a number of auxiliary components aid specific elements of either the training or generation process. 

The first set of additional components focuses on the latent space. As in standard VAE approaches, the latent space of the autoencoder must be regularized towards a known distribution, for which we make the standard choice of a Standard Normal distribution $\mathcal{N}(0,1)$. Two terms are included in the loss to enforce this latent regularisation constraint. The first is a Kullback-Leibler divergence (KLD) term, defined as %$\mathcal{L}_{KLD}$, defined as

\begin{equation}
    \mathcal{L}_{KLD}  = D_{KL}(N_E(x) || \mathcal{N}(0,1)),
\end{equation}
%\hfill\\

where $D_{KL}$ is the Kullback–Leibler divergence. It is defined between two probability distributions $P$ and $Q$, with probability densities $p$ and $q$ respectively, as 

\begin{equation} \label{KLD}
    D_{KL}(P || Q) =  \int p(x)\log\Bigg(\frac{p(x)}{q(x)}\Bigg) dx.
\end{equation}
%\hfill\\

\noindent The second latent regularisation contribution is a Maximum Mean Discrepancy (MMD)~\cite{MMD_base} term %$\mathcal{L}_{MMD}$

\begin{equation}
    \mathcal{L}_{MMD} = MMD(N_E(x), \mathcal{N}(0,1)),
\end{equation}

between the latent space and a Standard Normal distribution, defined as

\begin{equation}
    MMD^{2}(P,Q) = \mathbb{E} [k(x,x')] + \mathbb{E} [k(y,y')] - 2 \mathbb{E} [k(x,y)].
\end{equation}
%\hfill\\

where $x$, $x'$ and $y$, $y'$ are independent samples drawn from distributions $P$ and $Q$ respectively, $k$ is a positive definite function and $\mathbb{E}$ denotes the expectation. These are complemented by a loss contribution from a Wasserstein-GAN-like latent critic $C_{L}$, trained to distinguish between the latent space and a Standard Normal distribution, with loss given by %$\mathcal{L}_{Critic_{L}}$

\begin{equation}
 \mathcal{L}_{Critic_{L}} =\mathbb{E}[C_{L}(N_E(x))] .
\end{equation}
%\hfill\\

The approach to sampling from this latent space will be described in Section \ref{sec:Sampling}. 

The second role is fulfilled by a second Wasserstein-GAN-like reconstruction critic $C$, with loss contribution given by %$\mathcal{L}_{Critic}$. 

\begin{equation}
\mathcal{L}_{Critic} = \mathbb{E}[C(N_D(E(x)))].
\end{equation}

This critic serves a dual purpose --- it not only aids image reconstruction from the latent space by comparing input and output shower images, but also provides feedback as to whether showers look realistic or not. Note that following the developments in~\cite{HBFS},  the latent and reconstruction critics illustrated in Figure \ref{fig:BIBAE_PP_arch} actually represent two identical network architectures with independent weights. One network is trained continuously, while the other has its weights and optimizer reset after each epoch. This helps to prevent artefacts induced by the continuously trained critic becoming desensitised to particular data features, while retaining the learnt behaviour that is lost when resetting the critic. This was originally introduced in~\cite{HBFS} to deal with the additional sparsity present in hadronic showers, however we find that this also benefits the network performance for photon shower generation in the larger grid size studied here.

The combination of each of these elements results in a total loss given by
%\begin{equation}
%    \begin{gathered}
\begin{dmath}
        \mathcal{L}_{BIBA-AE} = \beta_{KLD} \cdot \mathcal{L}_{KLD} + \beta_{MMD} \cdot \mathcal{L}_{MMD} \\
                                + \beta_{Critic_{L}} \cdot \mathcal{L}_{Critic_{L}} + \beta_{Critic} \cdot  \mathcal{L}_{Critic},
\end{dmath}
%    \end{gathered}
%\end{equation}

\noindent with each term being controlled by an independent hyperparameter $\beta_{i}$~\cite{HBFS}.

The final component of the architecture is a separate Post Processor Network. This network consists of a series of kernel size one convolutions, which biases the network towards the refinement of individual voxel energies, rather than their creation or removal. By incorporating loss terms based around the Mean Squared Error (MSE) and the Sorted Kernel Maximum Mean Discrepancy (SK-MMD), as well as a loss term based on comparisons averaged over a batch of showers, the description of the hit energy spectrum can be significantly improved~\cite{GettingHigh}. This network is trained in a second step, once the training of the networks in the main pre-trained BIB-AE setup is complete and can be frozen, in order to enhance the stability of the Post Processor training~\cite{HBFS}.

% Mention mini-batch discrimination here!

Importantly, for the model to be useful as a simulator for physics it must be able to give a physically meaningful detector response that is dependent on the properties of the incident particle. For this reason, all of the networks in the BIB-AE architecture (with the exception of the latent critic) are conditioned on both the energy and polar angle of the incident particle. The Post Processor Network is additionally conditioned on the total visible energy each shower deposits in the calorimeter. The networks are built around combinations of 3D convolutions and fully connected layers, using the \textsc{ADAM} optimizer~\cite{adam} with an exponential learning rate decay. Minibatch discrimination is applied throughout the training. Prior to feeding the training data to the network, a threshold is applied to map hits below $1\times10^{-4}$ MeV to zero. The BIB-AE architecture was trained for a total of 50 epochs, after which the model was frozen and the Post Processor trained for a further 53 epochs. To select the best performing epochs, a similar approach to that described in~\cite{decoding_photons} was followed.  The selection was therefore based on a bin-wise area difference between distributions of key physics observables (see Section \ref{sec:SimResults}) for BIB-AE and \textsc{Geant4} generated showers. Particular weight was given to the visible energy and angular response, and a selection made to find the best performing base BIB-AE and Post Processing epoch.

\subsection{Sampling Strategy} \label{sec:Sampling}
Thus far, our description of the BIB-AE implementation has focused only on the steps necessary to train the model. However, if we are to use the model as a simulator, we must be able to draw samples from the learnt latent representation $z$. It is for this reason that generative autoencoder approaches introduce structure to the latent space via regularisation. This regularisation, however, comes at a cost -- the more we drive our latent space to have a known structure, the less information we can encode. Inspired by the buffer VAE approach~\cite{Buffer_VAE}, previous BIB-AE studies~\cite{decoding_photons}~\cite{HBFS} have applied density estimation to permit an improved latent space sampling. The advantage of this method is that it reduces the regularization constraint, and allows for an increased focus on information retention. Kernel Density Estimation~\cite{KDE} was previously used for this purpose in~\cite{decoding_photons}~\cite{HBFS}, providing an accurate model of the latent space variables and their correlations, and resulting in a performant simulator. However, since both of these studies were restricted to conditioning on only the energy of the incident particle, they made use of a rejection sampling method to generate samples of a specific energy. While this sampling method is sufficiently fast for single parameter conditioning, it becomes a major bottleneck for multi-parameter conditioning. A new conditional density estimation strategy is therefore required. 

A class of models that are well suited to this task are so called \textit{Normalizing Flows}~\cite{VIwithNF}~\cite{VIwithIAF}. These models aim to learn a bijective mapping $X = g(Z)$ with inverse $f := g^{-1}$ between a simple \textit{base} random variable $z\in{Z}$ with a known and tractable probability density function $p_{z}(z)$, and a random variable $x\in{X}$ with some unknown probability density function $p_{x}(x)$. Under the change of variables formula, $p_x(x)$ can then be computed from $p_{z}(z)$ via

\begin{equation}
    p_{x}(x) = p_{z}(f(x)) \Bigg \lvert det \frac{\partial f(x)}{\partial x} \Bigg \rvert = p_{z}(z) \Bigg \lvert det \frac{\partial g(z)}{\partial z} \Bigg \rvert^{-1}.
\end{equation}
\hfill\\

Our Normalizing Flow model is implemented using the \textsc{PYRO}~\cite{bingham2019pyro} deep probabilistic programming library. The model itself is a hybrid, consisting of $8$ blocks each with $7$ coupling layers. Of these coupling layers, $6$ are based on affine transformations~\cite{RealNVP} and one is based on element-wise rational spline bijections of linear order~\cite{NSplineFlows}~\cite{GenerativeRationalSplines}. Each layer is conditioned on a two dimensional context containing the energy and angular labels, which are pre-scaled by dividing by values of $100$ and $\frac{\pi}{2}$ respectively. To train the model, the 500k shower samples are encoded with the pre-trained BIB-AE model, resulting in 24 latent variables for each shower. Additionally, the corresponding energy sum of each shower in MeV scaled by a factor of $10^{4}$ is appended, resulting in a 25-dimensional training distribution. The inclusion of the energy sum of the shower here not only provides the additional conditioning label to the Post Processor Network (see Section~\ref{sec:BIBAE}) during sampling, but also permits a per-shower re-scaling of the visible energy sum in a similar manner to~\cite{Krause:2021Flow1} (see Figure~\ref{fig:BIBAE_PP_arch}). The Normalizing Flow model was trained for a total of $200$ epochs, and the best performing epoch selected based on the loss. This parameter configuration of the model was then used for subsequent latent space sampling to feed into the pre-trained BIB-AE during inference.

% Some reasoning here for this choice? E.g. we found this choice of architecture to be suitably expressive for the low dimensional density estimation task to which it is applied, while crucially maintaining low training and sampling times 

\section{Results}\label{sec:Results}
In order to benchmark the physics performance of the BIB-AE generator, we compare statistical distributions of key physics observables produced by the model to those of \textsc{Geant4} in Section \ref{sec:SimResults}. This comparison is broken down into two parts.

Firstly, we look at \textit{simulation level}. Here the direct output of the two simulators is used, providing an indication of how well the BIB-AE model has learnt from the training data. This has been the typical means of comparison in the vast majority of prior work (e.g.~\cite{GettingHigh}~\cite{Khattak:2021ndw}), and while it provides a useful and interesting comparison, it does not provide a complete picture.

For this reason, we additionally perform a comparison of observables at \textit{reconstruction level}, which are ultimately the quantities that will be used in physics analyses. Since the reconstruction procedure (see Section~\ref{sec:Reco}) typically relies on applying a series of sophisticated topological clustering algorithms, it is by no means clear a priori which attributes of the high dimensional data space are important for a model to capture. This was demonstrated previously in~\cite{HBFS}. Here we go beyond that study, which looked only at the effect of reconstruction on linearity and resolution, by providing a systematic study of the effect of reconstruction on a large number of observables.

We conclude the results by reporting the computational performance of the BIB-AE model during inference in comparison to \textsc{Geant4} in Section~\ref{sec:COmpPerf}.

\subsection{Physics Performance}\label{sec:SimResults}
For each observable, we begin by studying the performance of the BIB-AE in comparison to \textsc{Geant} before reconstruction, but with a simple calibration factor to account for the two sampling fractions (as described in~\ref{sec:Dataset}).

A cut is applied to cells with an energy deposition below $0.07875$ MeV, which corresponds to cell energies that are less than half of the most probable energy loss of a minimum ionising particle (MIP). This emulates the situation in a real calorimeter, since this region lies below the noise floor.

Subsequently, the effect of reconstruction (see Section~\ref{sec:Reco}) on each observable is studied. For this purpose, a selection criteria is placed on the data, such that only events containing one PFO are used for the performance evaluation.

\begin{figure*}[h]
    \centering
    \includegraphics[width=0.32\textwidth]{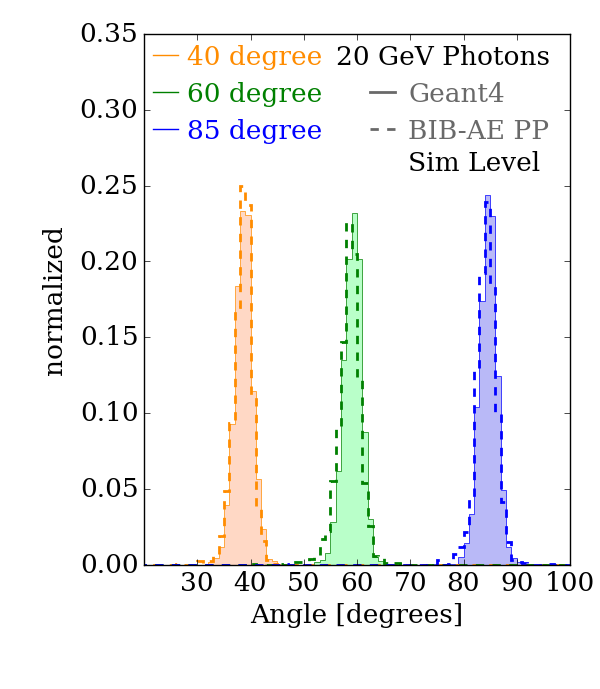}%{Plots_colours/Angles_3ofakind_20GeV.png}
    \includegraphics[width=0.32\textwidth]{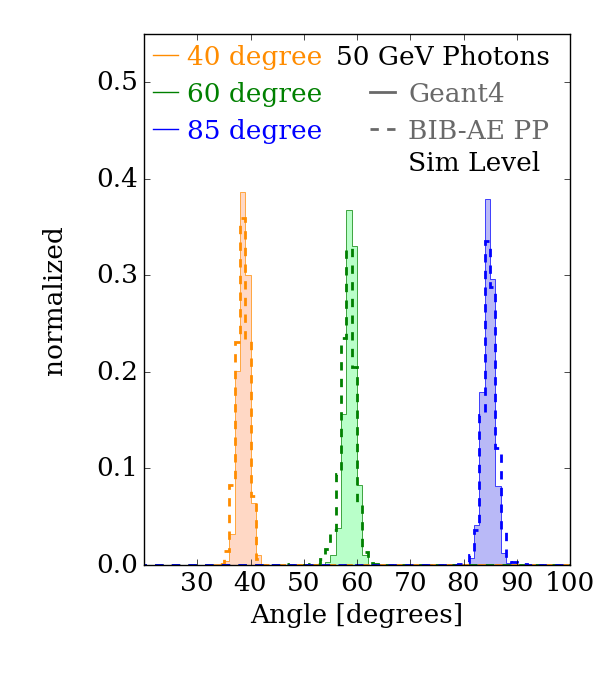}%{Plots_colours/Angles_3ofakind_50GeV.png}
    \includegraphics[width=0.32\textwidth]{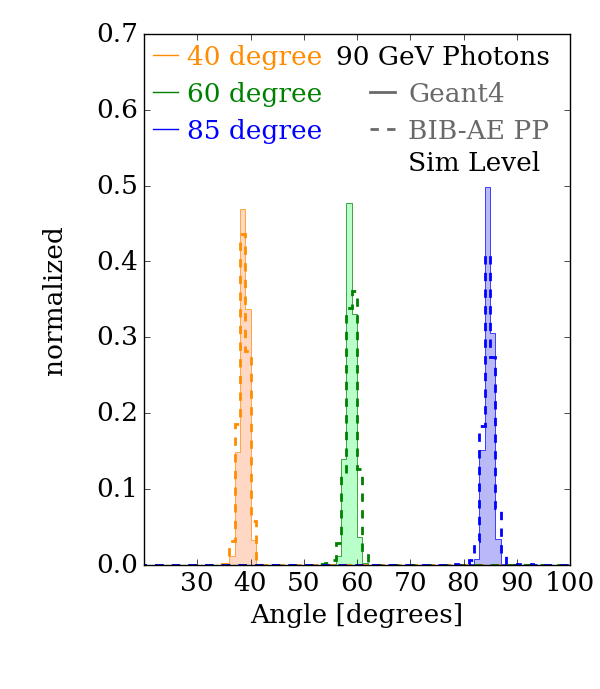}%{Plots_colours/Angles_3ofakind_90GeV.png}
    \caption{Angular response at simulation level for both \textsc{Geant4} (filled histograms), and BIB-AE generated photon showers (dashed, unfilled histograms). Distributions are shown for each incident energy of $20$ GeV (left), $50$ GeV (center) and $90$ GeV (right), and for each incident angle of $40$ degrees (orange), $60$ degrees (green) and $85$ degrees (blue).
    }
    \label{fig:Sim_Angles}
    \hspace{0.5cm}
\end{figure*}

As a first step to quantify the quality of the BIB-AE simulator, we study the conditioning performance by considering observables that are highly correlated with the angle and energy of the incident photon.

\subsubsection{Angular Response}\label{sec:Ang}
\hfill\\
The angular response is characterised by finding the principal axis of the shower with a principle component analysis (PCA). The resulting distributions for the reconstructed angles in degrees for $20$ GeV (left), $50$ GeV (center) and $90$ GeV (right) showers are shown in Figure \ref{fig:Sim_Angles}, comparing the \textsc{Geant4} test data with BIB-AE generated data. In this figure and throughout this work, we apply a consistent colour scheme to represent distributions corresponding to showers with a fixed incident angle of 40 degrees (orange), 60 degrees (green) and 85 degrees (blue). Across the range of fixed angle and energy showers considered, the angular distributions of the BIB-AE generated showers match their \textsc{Geant4} counterparts. The most noticeable discrepancy is a slight mismodelling of the very sharp peaks present in the \textsc{Geant4} distributions by the BIB-AE, that tends to become more apparent as the energy of the incident photon is increased.

\begin{figure*}[h]
    \centering
    \includegraphics[width=0.32\textwidth]{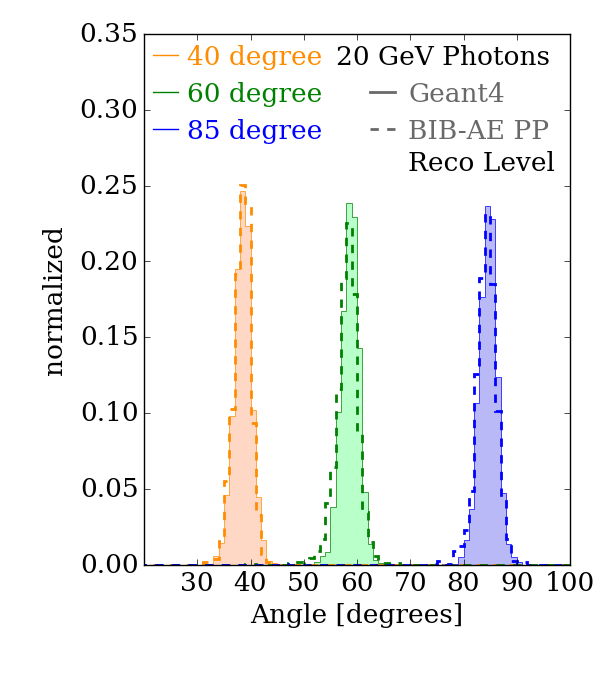}
    \includegraphics[width=0.32\textwidth]{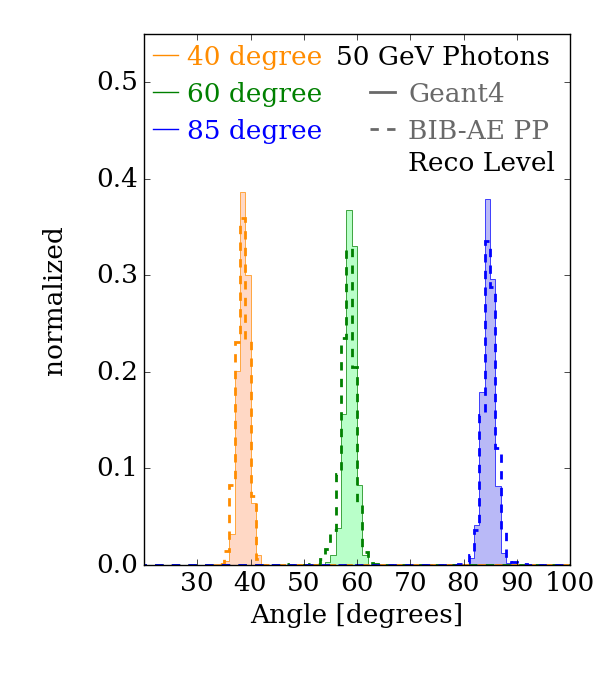}
    \includegraphics[width=0.32\textwidth]{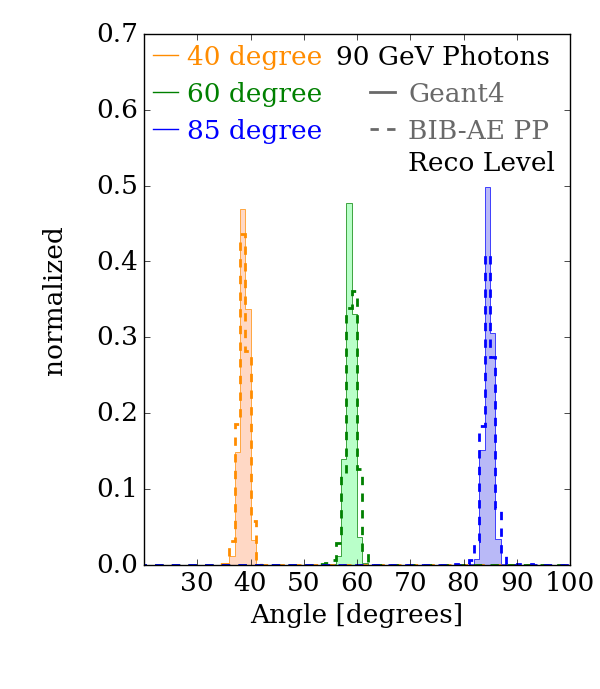}
    \caption{Reconstructed angular response for both \textsc{Geant4} (filled histograms), and BIB-AE generated photon showers (dashed, unfilled histograms). Distributions are shown for each incident energy of $20$ GeV (left), $50$ GeV (center) and $90$ GeV (right), and for each incident angle of $40$ degrees (orange), $60$ degrees (green) and $85$ degrees (blue).
    }
    \label{fig:Reco_Angles}
    \hspace{0.5cm}
\end{figure*}

The effect of reconstruction on the angular performance of the model can be seen in Figure~\ref{fig:Reco_Angles}. The reconstructed angle of each shower is obtained from the intrinsic direction of the clustered hits. Reconstruction has little visible effect on the angle, with the overall good agreement to \textsc{Geant4}, and the slight mismodelling of the sharp peaks at higher energies, being retained.

\begin{figure*}[h]
    \centering
    \includegraphics[width=0.45\textwidth]{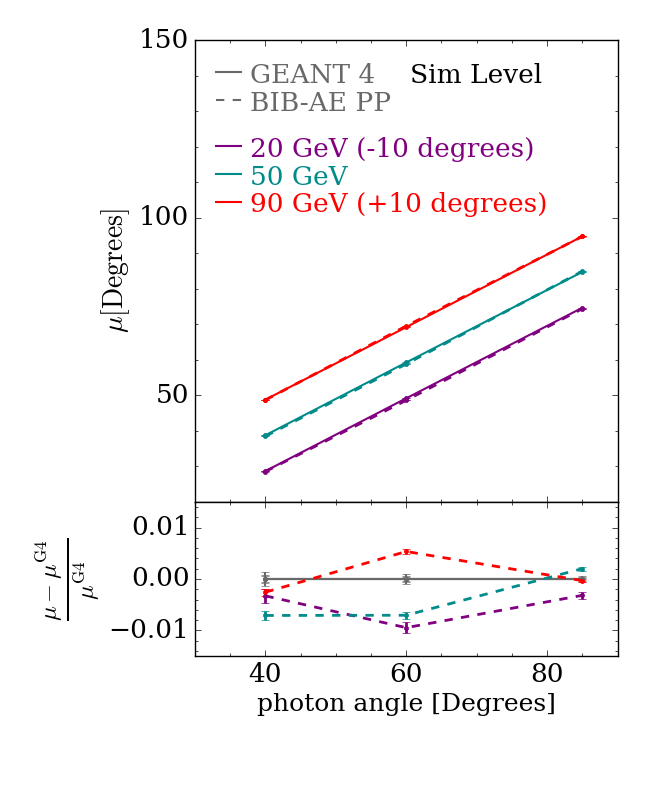}
    \includegraphics[width=0.45\textwidth]{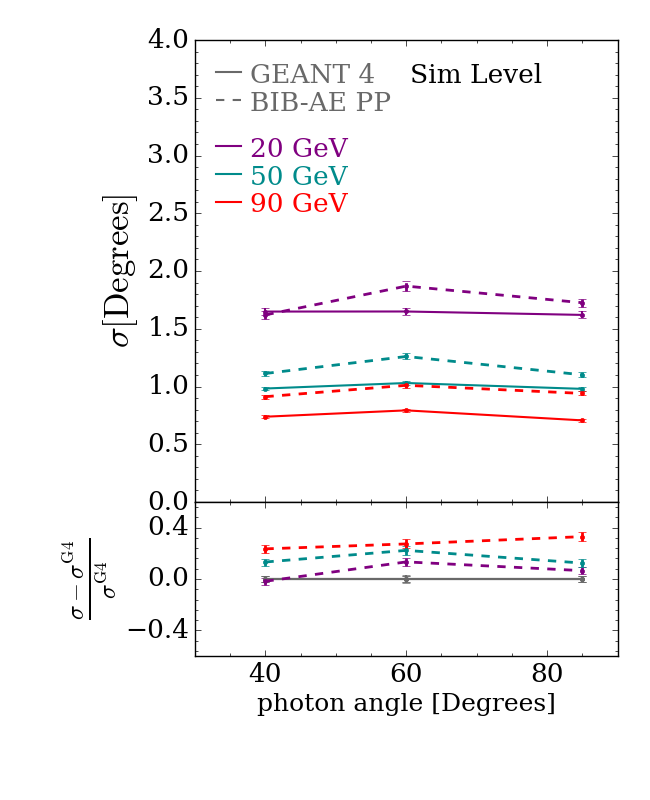}
    \caption{Simulation level angular linearity (left) and width (right) for both \textsc{Geant4} and BIB-AE generated showers. Curves are shown for each of the fixed incident energies of $20$ GeV, $50$ GeV and $90$ GeV, which are coloured purple, dark cyan and red respectively. In the angular linearity plot on the left, the means for showers with energies of $20$ GeV and $90$ GeV are shifted by constant values of $-10$ degrees and $+10$ degrees respectively for visual purposes. The sub-panels in each figure show the relative deviation of the BIB-AE angular responses from their \textsc{Geant4} equivalents.
    }
    \label{fig:Sim_AngleLinRes}
    \hspace{0.5cm}
\end{figure*}

The angular distributions are characterised in more detail in Figure~\ref{fig:Sim_AngleLinRes} through Gaussian fits to both the \textsc{Geant4} and BIB-AE distributions. The mean $\mu$ and standard deviation $\sigma$ of the fits are then extracted, and the means (left) and standard deviations (right) plotted as functions of the incident particle angles in order to obtain the effective angular linearity and widths for each fixed incident energy of $20$ GeV, $50$ GeV and $90$ GeV. The means of the angular distributions are very well reproduced by the BIB-AE, with the maximum deviations from the \textsc{Geant4} values reaching only to the $1$ \% level. Turning to the resolution plot in the right of the figure, the performance of the BIB-AE appears to degrade with increasing energy. The angular response for $20$ GeV showers at $40$ degrees shows excellent agreement, whereas a deviation of almost $40$\% is observed for $90$ GeV showers at $85$ degrees.

\begin{figure*}[h]
    \centering
    \includegraphics[width=0.45\textwidth]{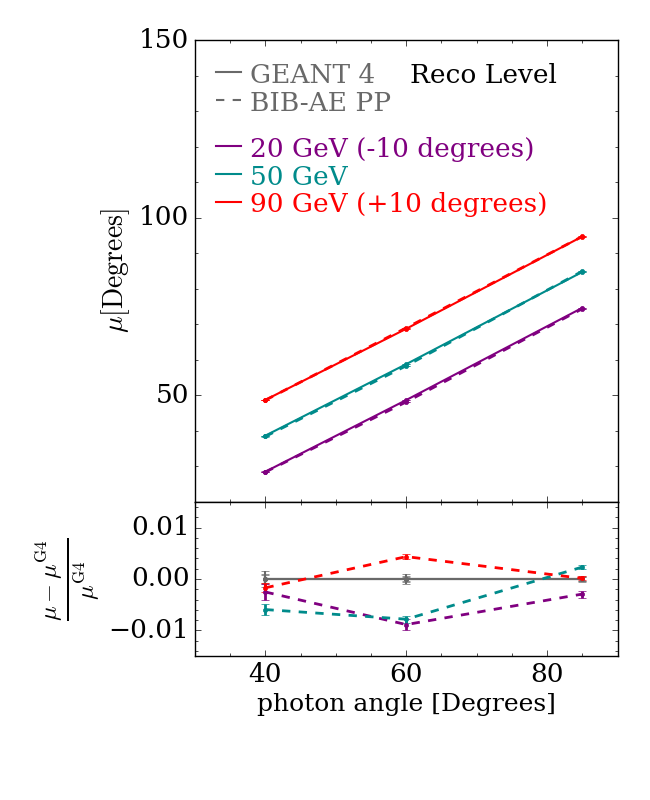}
    \includegraphics[width=0.45\textwidth]{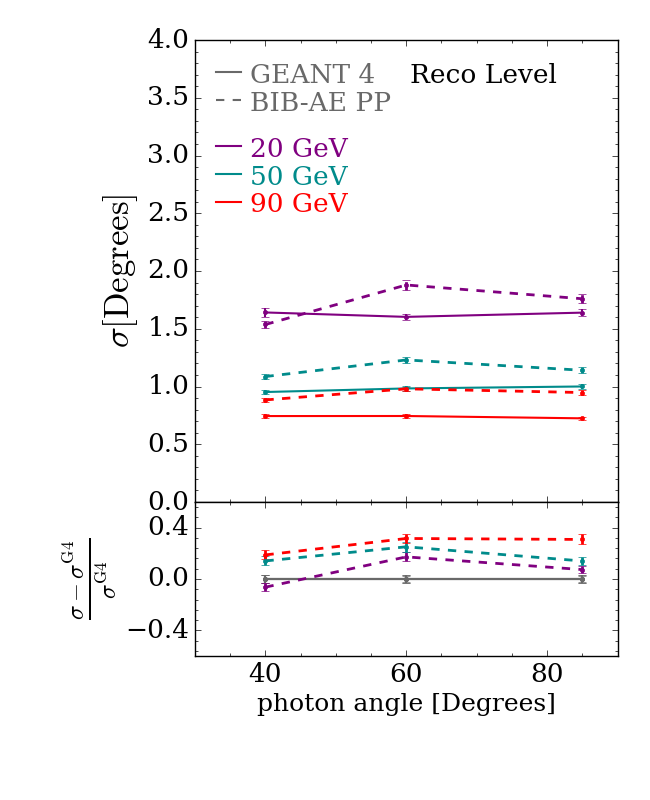}
    \caption{Reconstructed angular linearity (left) and width (right) for both \textsc{Geant4} and BIB-AE generated showers. Curves are shown for each of the fixed incident energies of $20$ GeV, $50$ GeV and $90$ GeV, which are coloured purple, dark cyan and red respectively. In the angular linearity plot on the left, the means for showers with energies of $20$ GeV and $90$ GeV are shifted by constant values of $-10$ degrees and $+10$ degrees respectively for visual purposes. The sub-panels in each figure show the relative deviation of the BIB-AE angular responses from their \textsc{Geant4} equivalents.
    }
    \label{fig:Reco_AngleLinRes}
    \hspace{0.5cm}
\end{figure*}

The effects of reconstruction on the angular linearity and width can be seen in Figure~\ref{fig:Reco_AngleLinRes}. This confirms that reconstruction has a minimal impact on the angular performance of the BIB-AE, with only minor changes in the mean and widths of the angular distributions being introduced by the clustering procedure in reconstruction.

\subsubsection{Energy Response}\label{sec:Energy}
\begin{figure*}[h]
    \centering
    \includegraphics[width=0.3\textwidth]{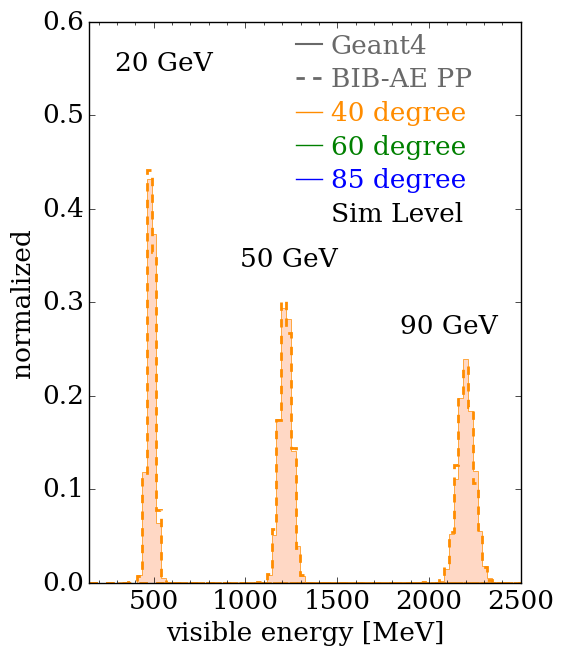}%Plots_colours/single_energy_3ofakind_40degrees.png}
    \includegraphics[width=0.3\textwidth]{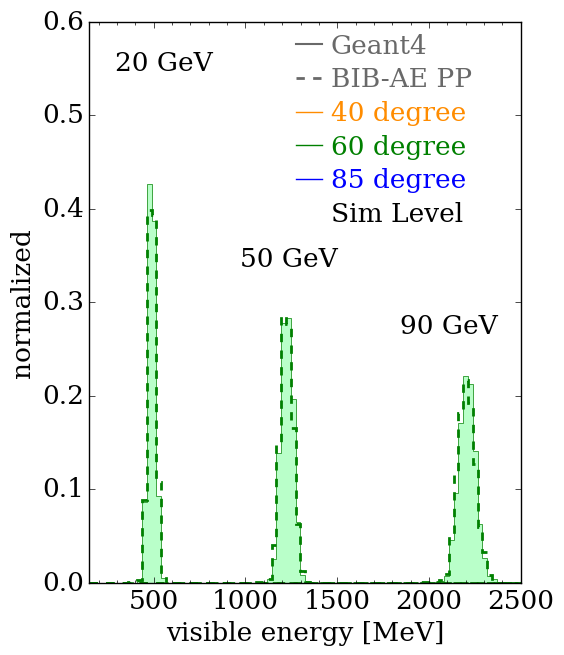}%Plots_colours/single_energy_3ofakind_60degrees.png}
    \includegraphics[width=0.3\textwidth]{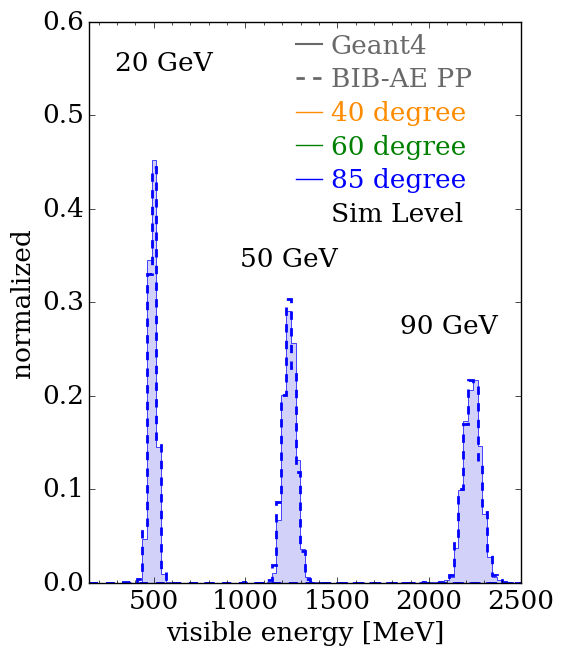}%Plots_colours/single_energy_3ofakind_85degrees.png}
    \caption{Visible energy deposited in the calorimeter at simulation level for both \textsc{Geant4} and BIB-AE generated showers. The distributions are grouped according to incident photon angles of $40$ degrees (left, orange), $60$ degrees (center, green) and $85$ degrees (right, blue).
    }
    \label{fig:Sim_EnergySum}
    \hspace{0.5cm}
\end{figure*}

\begin{figure*}[h]
    \centering
    \includegraphics[width=0.45\textwidth]{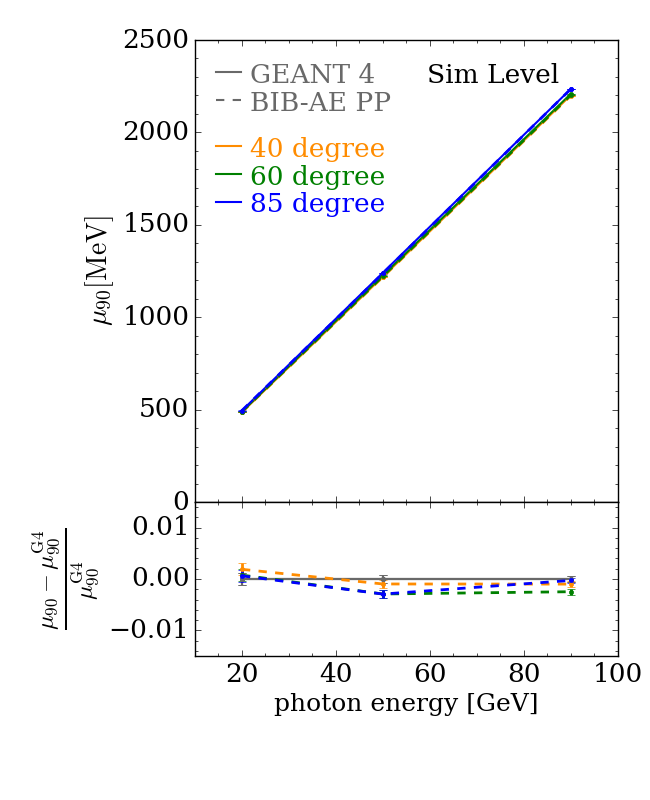}%Plots_colours/Energy_linearity.png}
    \includegraphics[width=0.45\textwidth]{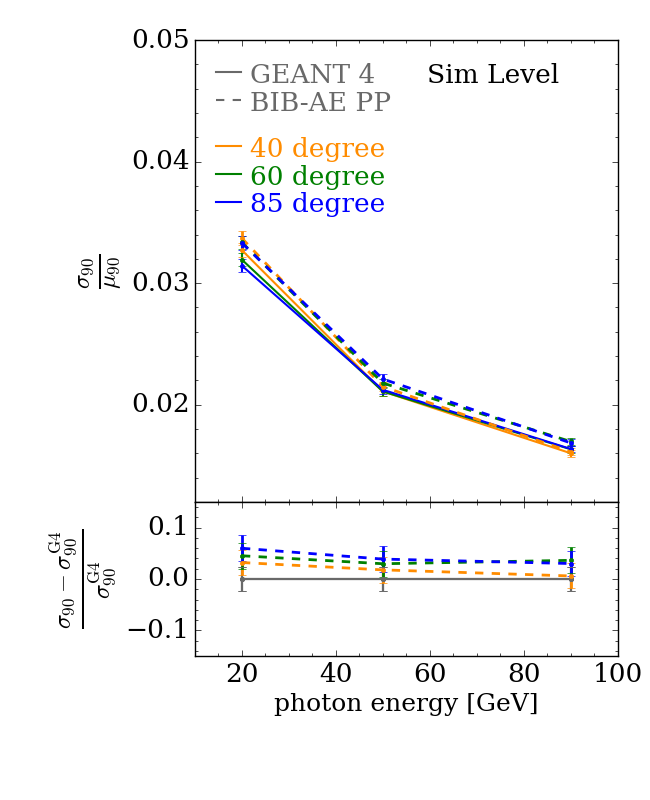}%Plots_colours/Energy_resolution.png}
    \caption{Energy linearity (left) and resolution (right) at simulation level for for both \textsc{Geant4} and BIB-AE generated showers. The curves are grouped according to the three incident angles of $40$ degrees (orange), $60$ degrees (green) and $85$ degrees (blue). The sub-panels in each figure show the relative deviation of the BIB-AE visible energy responses from their \textsc{Geant4} equivalents.
    }
    \label{fig:Sim_EnLinRes}
    \hspace{0.5cm}
\end{figure*}

\hfill\\
To study the energy conditioning performance of the BIB-AE, the total energy deposited in the active sections of the calorimeter is determined and shown in Figure \ref{fig:Sim_EnergySum}. The energy sums of the BIB-AE distributions match their \textsc{Geant4} equivalents across the board, with only minor deviations. This is a direct result of the re-scaling procedure using the per-shower energy sums generated by the Normalising Flow (see Section~\ref{sec:Sampling}). The energy sum distributions are characterised in further detail in Figure \ref{fig:Sim_EnLinRes}, with the mean ($\mu_{90}$) and rms ($\sigma_{90}$) of the central $90$\% of the distributions being calculated. These are shown in the Figure as the linearity and resolution as a function of the incident particle energy, for each of the various incident angles. The linearity (left) is particularly well described, especially for photons with a low incidence angle, with the maximum deviations from the \textsc{Geant4} value restricted to well below the $1$\% level. While the resolution (right) exhibits larger deviations, it is also well described, with deviations being restricted to below the $10$\% level.

\begin{figure*}[h]
    \centering
    \includegraphics[width=0.3\textwidth]{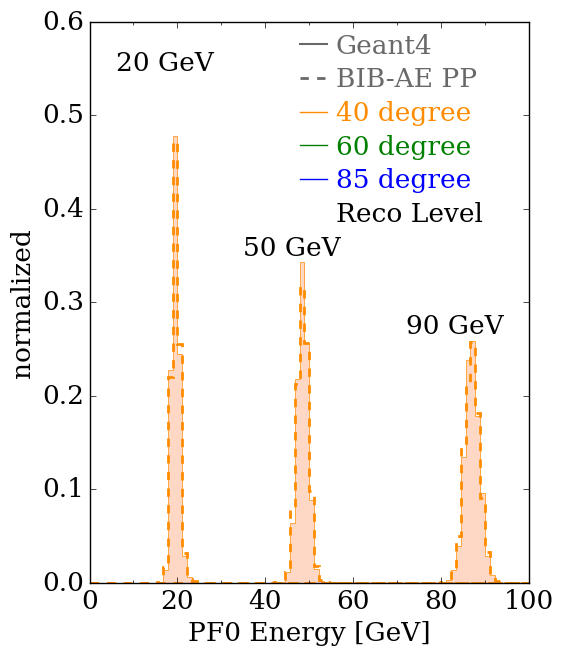}
    \includegraphics[width=0.3\textwidth]{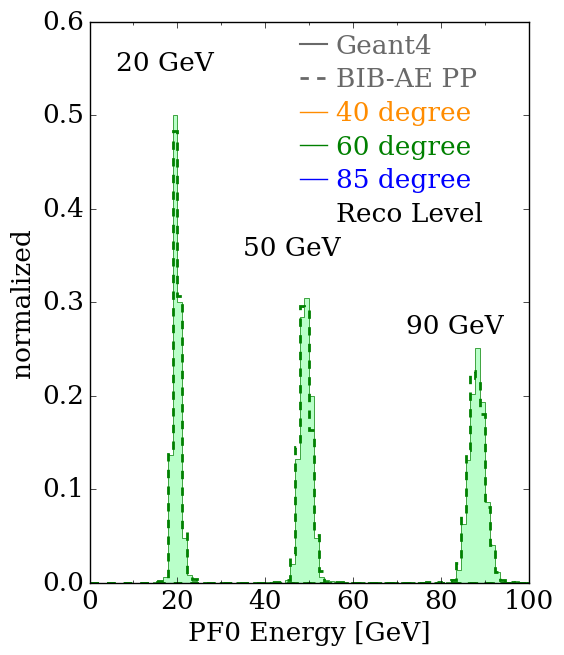}
    \includegraphics[width=0.3\textwidth]{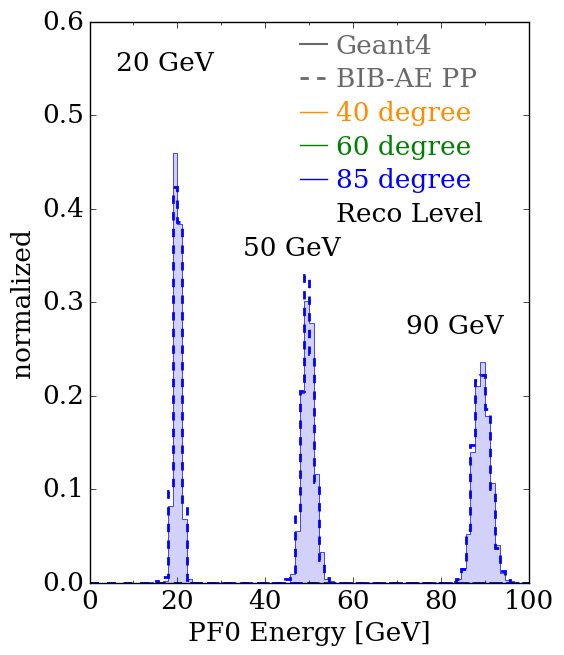}
    \caption{Reconstructed particle (PFO) energy for both \textsc{Geant4} and BIB-AE generated showers. The distributions are grouped according to incident photon angles of $40$ degrees (left, orange), $60$ degrees (center, green) and $85$ degrees (right, blue).
    }
    \label{fig:Reco_EnergySum}
    \hspace{0.5cm}
\end{figure*}

\begin{figure*}[h]
    \centering
    \includegraphics[width=0.45\textwidth]{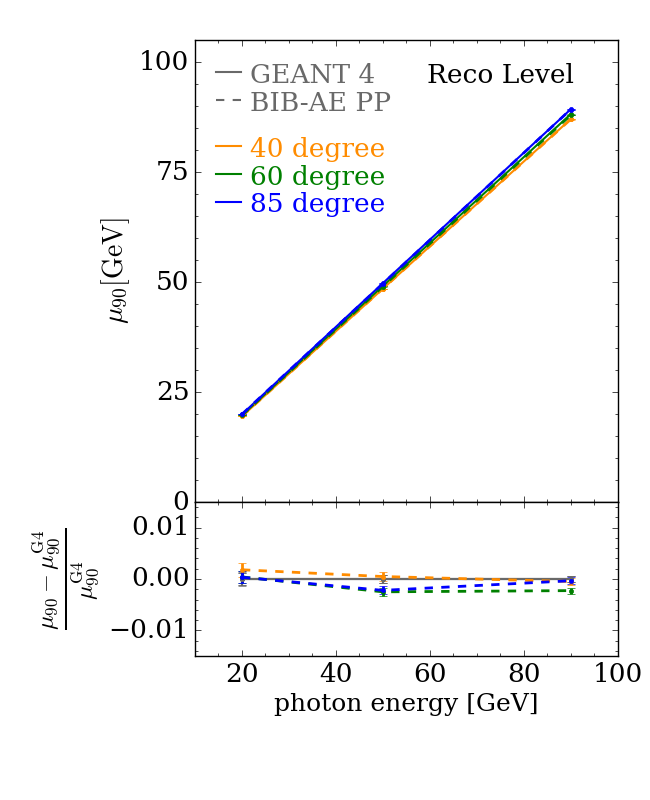}
    \includegraphics[width=0.45\textwidth]{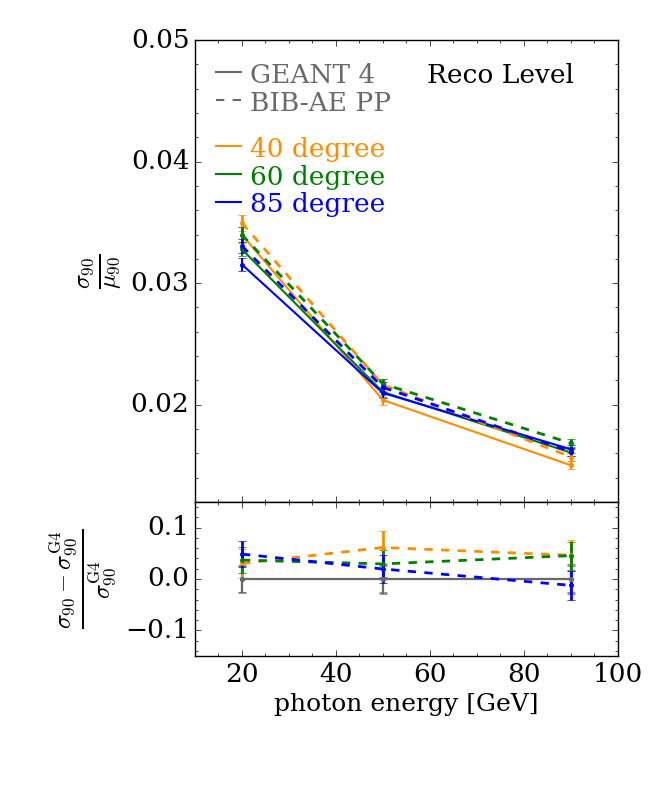}
    \caption{Reconstructed particle (PFO) energy linearity (left) and resolution (right) for both \textsc{Geant4} and BIB-AE generated showers. The curves are grouped according to the three incident angles of $40$ degrees (orange), $60$ degrees (green) and $85$ degrees (blue). The sub-panels in each figure show the relative deviation of the BIB-AE visible energy responses from their \textsc{Geant4} equivalents.
    }
    \label{fig:Reco_EnLinRes}
    \hspace{0.5cm}
\end{figure*}

The effect of reconstruction on the energy conditioning performance of the model can be seen in Figure~\ref{fig:Reco_EnergySum}. In this case, the distributions show the energy of the Particle Flow Object (PFO) reconstructed by PandoraPFA, which correspond to the incident particle energy that has been reconstructed. Again, an excellent agreement between the BIB-AE and \textsc{Geant4} distributions is observed. This is confirmed by the PFO linearity and resolution shown in Figure~\ref{fig:Reco_EnLinRes}. The effect of reconstruction on the linearity is minimal. While a slight effect on the resolution can be seen, these deviations are still restricted to below the $10$\% level across all the test points.

\subsubsection{Cell Energy Spectrum}\label{sec:CellE}
\hfill\\
We now turn our attention to the cell energy spectra of the showers. The results are summarised in Table~\ref{table:Sim_HitE}, by calculating the Jensen-Shannon Distance (JSD) between the \textsc{Geant4} and the BIB-AE distributions for each combination of energy and angle. The JSD between two probability distributions $P$ and $Q$, with probability densities $p$ and $q$ respectively, is the square root of the Jensen-Shannon divergence

\begin{equation}
    \JSD(P || Q) = \sqrt{\frac{D_{KL}(P || M) + D_{KL}(Q || M)}{2}},
\end{equation}
\hfill\\
where $M = \frac{1}{2}(P + Q)$, and $D_{KL}$ is the Kullback–Leibler divergence, defined in equation~\ref{KLD} in Section~\ref{sec:BIBAE}.
%as

%\begin{equation}
%    D_{KL}(P || Q) =  \int p(x)\log\Bigg(\frac{p(x)}{q(x)}\Bigg) dx.
%\end{equation}
%\hfill\\

%\begin{center}
%\begin{tabular}{ll|SSS|SSS}%{@{}p{0.12\textwidth}*{4}{L{\dimexpr0.22\textwidth-2\tabcolsep\relax}}@{}}
%\toprule
%
%                                 &&& {Energy} \\
%                                 &&& (GeV)\\\midrule
%                                 &&& SIM &&& REC \\
%                          && 20  & 50  & 90                  & 20  & 50  & 90 \\\bottomrule
%Angle              & 40  &  0.029 & 0.024 & \bf{0.036}   &  \bf{0.031} & 0.022 & 0.029 \\
%(deg.)             & 60  &  0.026 & 0.021 & 0.032      &  0.027 & 0.018 & 0.025 \\
%                   & 85  &  0.028 & \bf{0.018} & 0.029 &  0.029 & \bf{0.015} & 0.021\\
%                    \bottomrule

%\end{tabular}
%\end{center}
%\vspace{15pt}
%\end{table*}

\begin{table*}[tbh]
%\sisetup{
%separate-uncertainty=true,
%table-format=4.3(5)
%}
\setlength{\tabcolsep}{12pt}
\centering

\caption{Table showing the Jensen Shannon Distance (JSD) between the Geant4 and BIB-AE results for the simulation and reconstruction level cell energy spectrum for each combination of fixed energy and angle. For the simulation level results, the reported JSD is only calculated in the region above the MIP cut.}\label{table:Sim_HitE}
\vspace{15pt}

\begin{center}
\let\mc\multicolumn
    \centering
    \begin{tabular}{ll|P{1.0cm}P{1.0cm}P{1.0cm}|P{1.0cm}P{1.0cm}P{1.0cm}}
    %\begin{tabular}{ll|SSS|SSS}%{@{}p{0.12\textwidth}*{4}{L{\dimexpr0.22\textwidth-2\tabcolsep\relax}}@{}}
        \toprule

                                 &&& \mc4c{Energy} \\
                                 &&& \mc4c{(GeV)}\\\midrule
                                 &&& SIM &&& REC \\
                          && 20  & 50  & 90                  & 20  & 50  & 90 \\\bottomrule
Angle              & 40  &  0.029 & 0.024 & \bf{0.036}   &  \bf{0.031} & 0.022 & 0.029 \\
(deg.)             & 60  &  0.026 & 0.021 & 0.032      &  0.027 & 0.018 & 0.025 \\
                   & 85  &  0.028 & \bf{0.018} & 0.029 &  0.029 & \bf{0.015} & 0.021\\
                    \bottomrule

        \end{tabular}

\end{center}
\vspace{15pt}
\end{table*}

\begin{figure*}[h]
    \centering
    \includegraphics[width=0.40\textwidth]{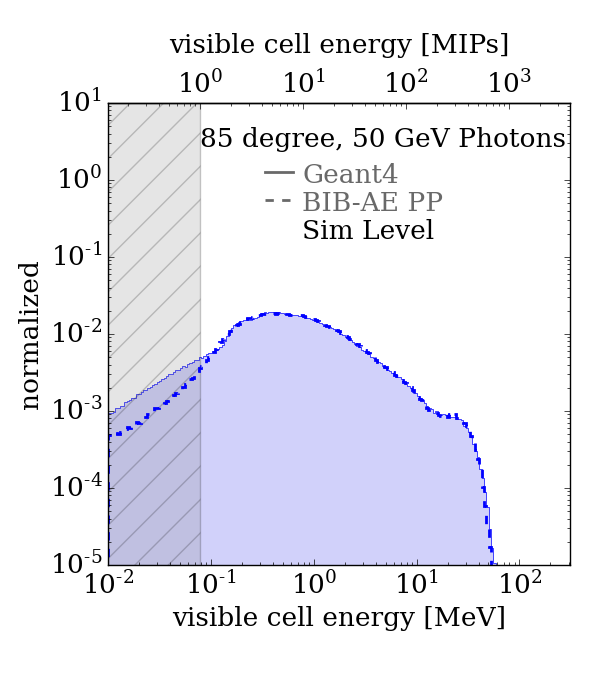}%Plots_colours/Hit_E_40degree_20GeV.png}
    \includegraphics[width=0.40\textwidth]{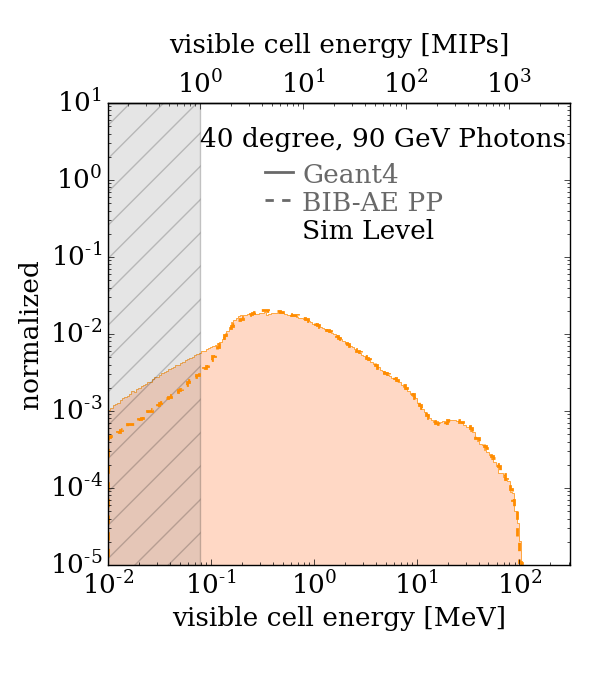}%Plots_colours/Hit_E_60degree_90GeV.png}
    \caption{Simulation level cell energy spectra for the best ($50$ GeV, $85$ degrees, left) and worst ($90$ GeV, $40$ degrees, right) incident angle and energy combinations. The grey hatched area indicates the region below half a MIP.}
    \label{fig:Sim_HitE}
\end{figure*}

\begin{figure*}[h]
    \centering
    \includegraphics[width=0.40\textwidth]{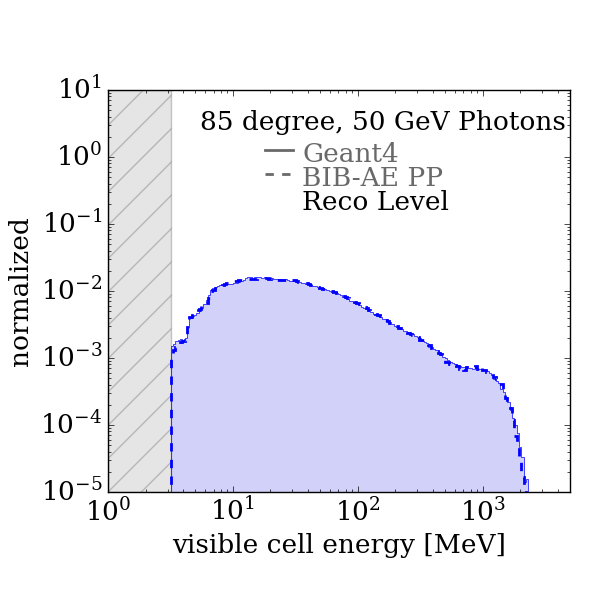}
    \includegraphics[width=0.40\textwidth]{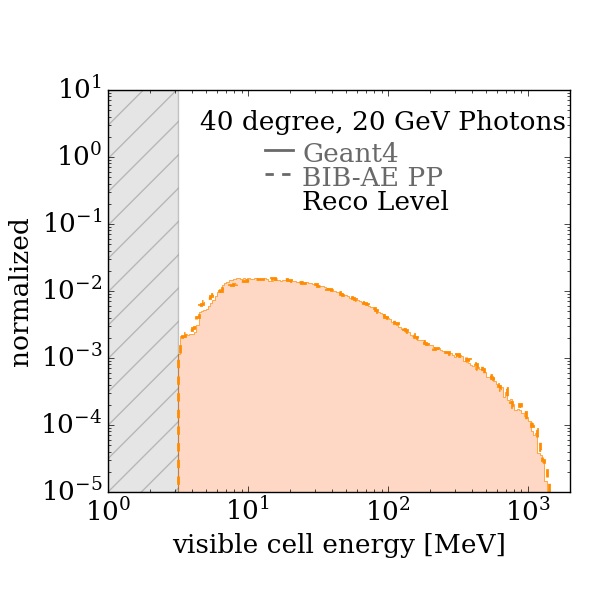}
    \caption{Reconstructed cell energy spectra for the best ($50$ GeV, $85$ degrees, left) and worst ($20$ GeV, $40$ degrees, right) incident angle and energy combinations. The grey hatched area indicates the region below half a MIP.}
    \label{fig:Reco_HitE}
\end{figure*}

For the simulation level results, the best ($50$ GeV, $85$ degree) and worst ($90$ GeV, $40$ degree) combinations are highlighted and presented in Figure~\ref{fig:Sim_HitE}. The grey hatched region to the left of these plots indicates the region below half a MIP that will be removed in a real calorimeter (typically referred to as the MIP cut). Therefore, for the simulation level results, the reported JSD is only calculated in the region above the MIP cut. Larger deviations from the \textsc{Geant4} ground truth are exhibited in the region around and below this cut-off. However, above the half MIP cut the variations in the shape of this spectrum with incident particle energy and angle tend to be accurately described. The key feature is the distinct peak that occurs in the spectrum around $1$ MIP, which is well reproduced by the BIB-AE thanks to its Post Processor Network. This was first achieved by the original BIB-AE architecture~\cite{GettingHigh}, and these results indicate that this capability can be extended to more general simulation scenarios where multiple conditioning parameters are necessary.

The resulting JSD values for the cell energy spectrum of each energy and angle combination after reconstruction are summarised in the right of Table~\ref{table:Sim_HitE}. The best ($50$ GeV, $85$ degrees) and worst (now $20$ GeV, $40$ degrees) are highlighted in bold, and the distributions shown in Figure~\ref{fig:Reco_HitE}. These distributions end at the half MIP threshold, as lower energy hits are discarded during reconstruction. After reconstruction, the MIP peak that was smooth at simulation level is now smeared out as a result of the two MIP calibration factors that are applied during the process. Overall, the BIB-AE is able to reproduce the cell energy very well after reconstruction, with only minor deviations being visible, even for the worst performing distribution at $20$ GeV, $40$ degrees.

\subsubsection{Number of Hits}\label{sec:Nhits}
\hfill\\
\begin{figure*}[h]
    \centering
    \includegraphics[width=0.3\textwidth]{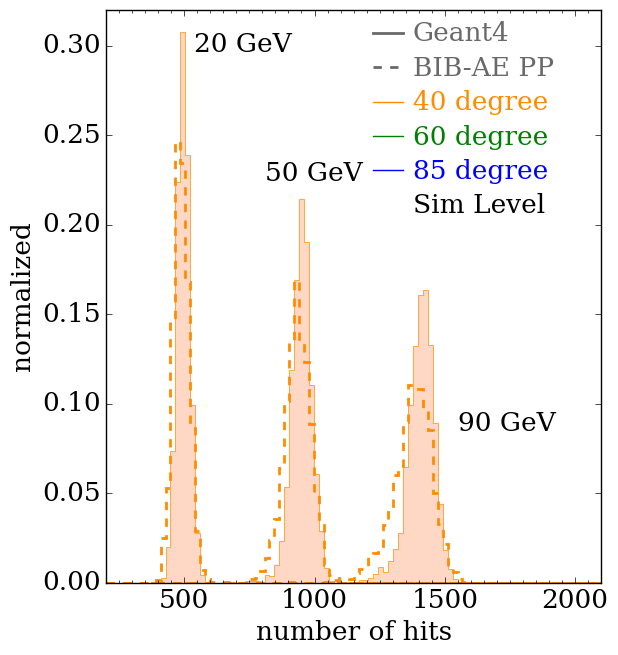}%Plots_colours/Occupancy_3ofakind_40degrees.png}
    \includegraphics[width=0.3\textwidth]{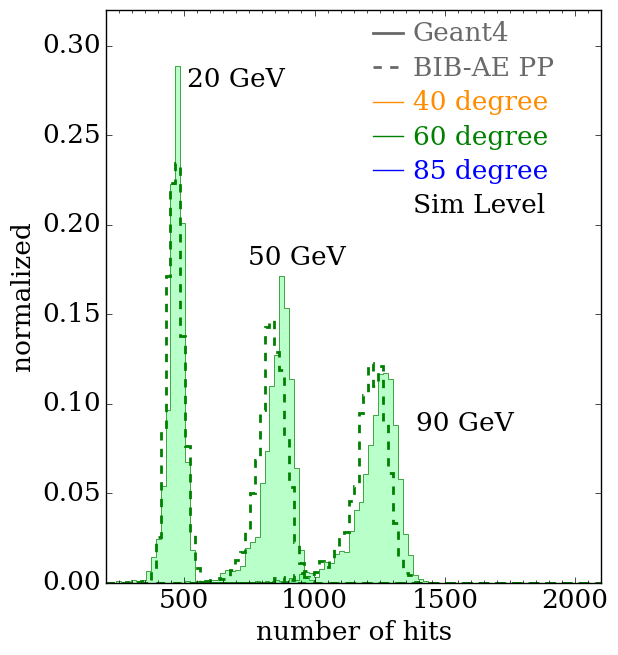}%Plots_colours/Occupancy_3ofakind_60degrees.png}
    \includegraphics[width=0.3\textwidth]{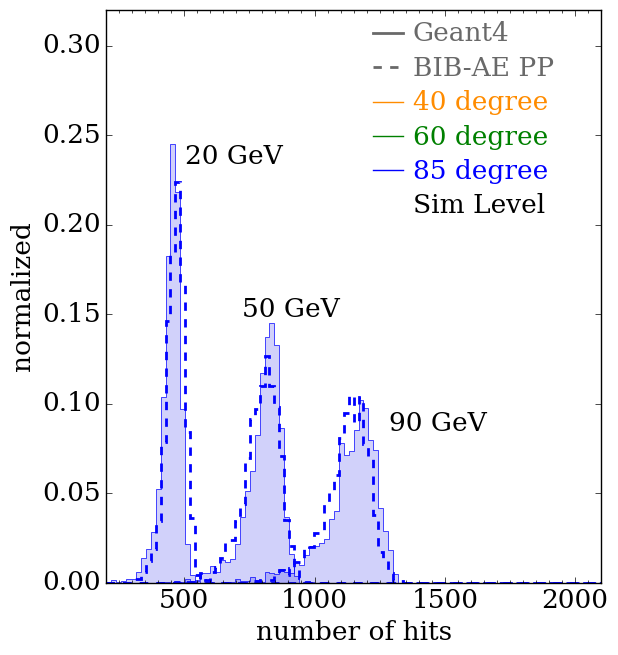}%Plots_colours/Occupancy_3ofakind_85degrees.png}
    \caption{Simulation level number of hits for both \textsc{Geant4} and BIB-AE generated showers. The distributions are grouped according to incident photon angles of $40$ degrees (left, orange), $60$ degrees (center, green) and $85$ degrees (right, blue).
    }
    \label{fig:Sim_NHits}
    \hspace{0.5cm}
\end{figure*}

\begin{figure*}[tbh]
    \centering
    \includegraphics[width=0.3\textwidth]{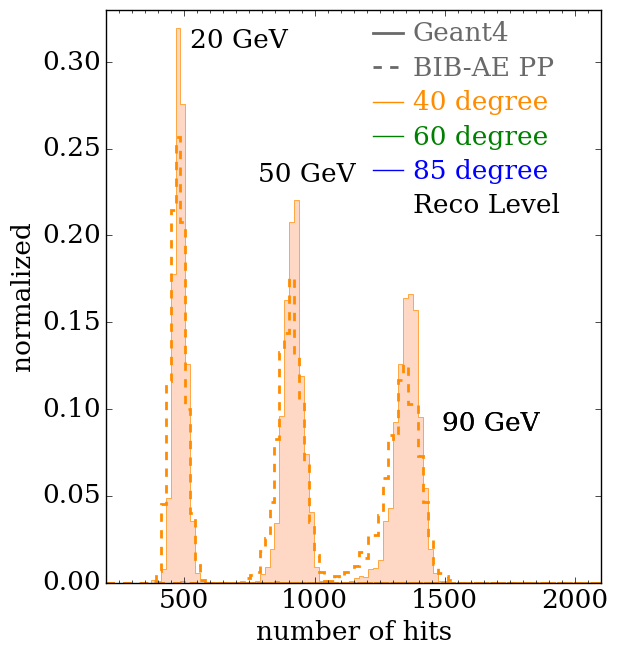}
    \includegraphics[width=0.3\textwidth]{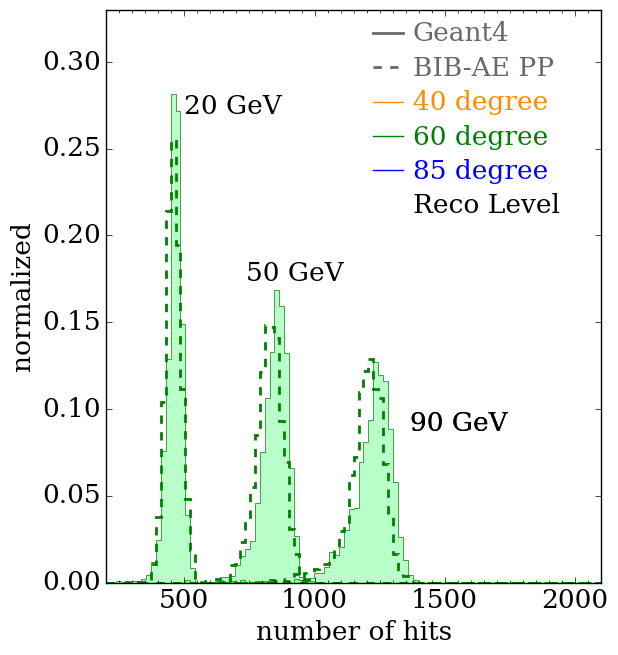}
    \includegraphics[width=0.3\textwidth]{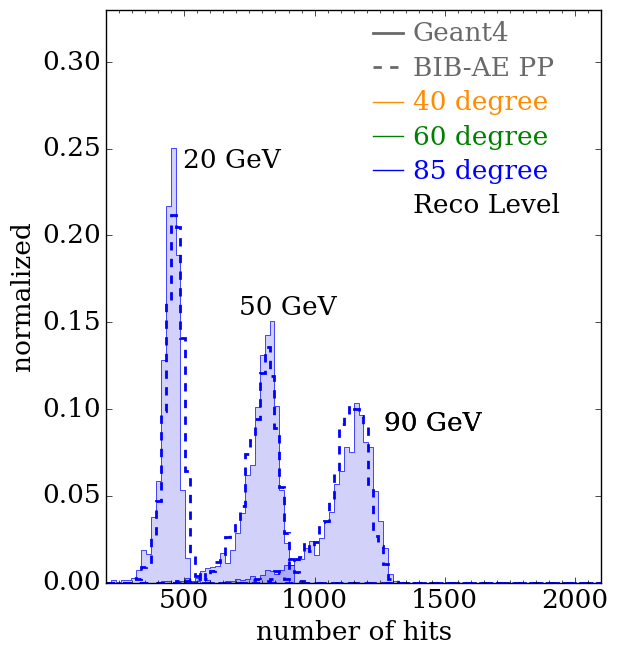}
    \caption{Reconstructed number of hits for both \textsc{Geant4} and BIB-AE generated showers. The distributions are grouped according to incident photon angles of $40$ degrees (left, orange), $60$ degrees (center, green) and $85$ degrees (right, blue).
    }
    \label{fig:Reco_NHits}
    \hspace{0.5cm}
\end{figure*}

Another distribution where the Post Processor Network can play a key role is in correctly producing the distribution of the number of hits with energy above the half MIP threshold shown in Figure~\ref{fig:Sim_NHits}. This distribution is strongly affected by how well the cell energy spectrum is modelled around the cut, and even slight discrepancies can significantly effect the resulting number of hits above the half MIP threshold. The differences that can be seen in these distributions, which tend to become more pronounced for higher energies where the number of hits increases, can therefore be linked to the more noticeable deviations in the cell energy spectrum around the cut.

The distributions of the number of hits after reconstruction are shown in Figure~\ref{fig:Reco_NHits}. These distributions do not show any significant shifts from the simulation level distributions, indicating that the vast majority of hits above the half MIP threshold are retained after reconstruction, as would be expected for electromagnetic showers.

%\begin{table*}[h!]
%\sisetup{
%separate-uncertainty=true,
%table-format=4.3(5)
%}
%\setlength{\tabcolsep}{12pt}
%\centering

%\caption{Table showing the Jensen Shannon Distance (JSD) between the Geant4 and BIB-AE results for the sim level center of gravity for each combination of fixed energy and angle.}\label{table:Sim_CoGZ}
%\vspace{15pt}
%\begin{center}
%\begin{tabular}{ll|SSS}%{@{}p{0.12\textwidth}*{4}{L{\dimexpr0.22\textwidth-2\tabcolsep\relax}}@{}}
%\toprule

%                                 &&& Energy \\
%                                 &&& (GeV)\\\midrule
%                          && 20  & 50  & 90\\\bottomrule
%Angle               & 40  &  0.066 & 0.062 & 0.072 \\
%(deg.)              & 60  &  0.062 & 0.059 & 0.058\\
%                    & 85  &  \bf{0.076} & \bf{0.057} & 0.070\\
%                    \bottomrule

%\end{tabular}
%\end{center}
%\vspace{15pt}
%\end{table*}

\subsubsection{Center of Gravity}\label{sec:CoGZ}
\hfill\\
\begin{table*}[h!]
%\sisetup{
%separate-uncertainty=true,
%table-format=4.3(5)
%}
\setlength{\tabcolsep}{12pt}
\centering

\caption{Table showing the Jensen Shannon Distance (JSD) between the Geant4 and BIB-AE results for the simulation and reconstruction level center of gravity for each combination of fixed energy and angle.}\label{table:Sim_CoGZ}
\vspace{15pt}
\begin{center}
\let\mc\multicolumn
    \centering
        \begin{tabular}{ll|P{1.0cm}P{1.0cm}P{1.0cm}|P{1.0cm}P{1.0cm}P{1.0cm}}
        %\begin{tabular}{ll|SSS|SSS}%{@{}p{0.12\textwidth}*{4}{L{\dimexpr0.22\textwidth-2\tabcolsep\relax}}@{}}
\toprule

                                 &&& \mc4c{Energy} \\
                                 &&& \mc4c{(GeV)}\\\midrule
                                 &&& SIM &&& REC \\
                          && 20  & 50  & 90                     & 20 & 50 & 90\\\bottomrule
Angle               & 40  &  0.066 & 0.062 & 0.072            & 0.081 & \bf{0.057} & 0.064\\
(deg.)              & 60  &  0.062 & 0.059 & 0.058            & 0.067 & 0.064 & 0.070\\
                    & 85  &  \bf{0.076} & \bf{0.057} & 0.070  & 0.079 & 0.067 & \bf{0.091}\\
                    \bottomrule

\end{tabular}
\end{center}
\vspace{15pt}
\end{table*}

\begin{figure*}[tbh]
    \centering
    \includegraphics[width=0.4\textwidth]{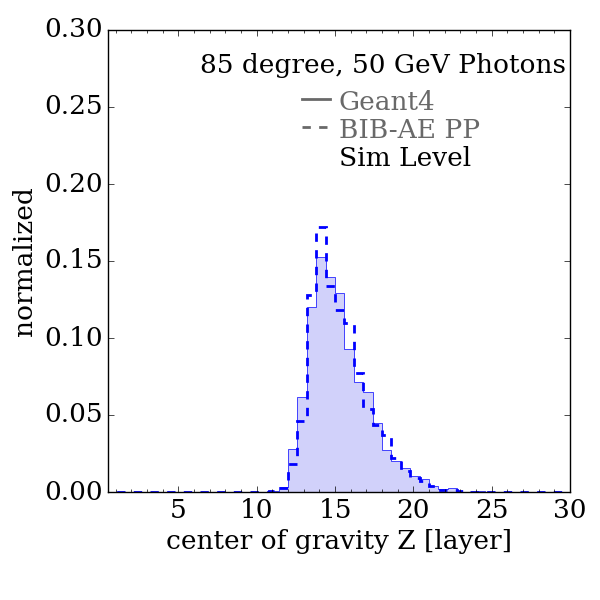}
    \includegraphics[width=0.4\textwidth]{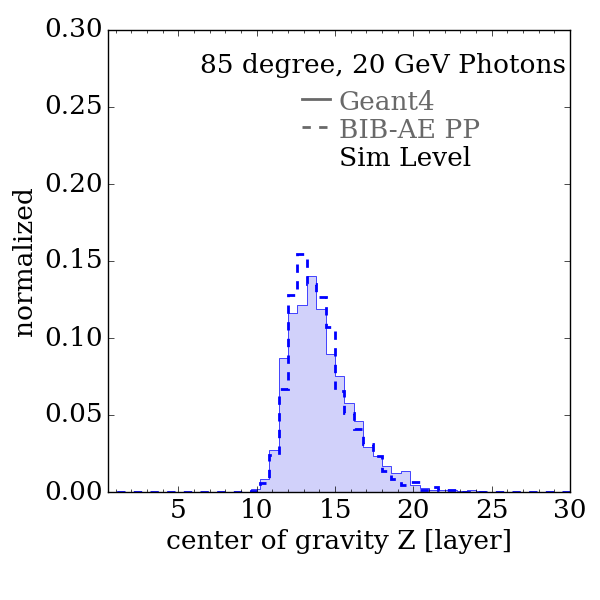}%Plots_colours/COG_Z_40degree_20GeV.png}
    \caption{Simulation level center of gravity distributions for the best ($50$ GeV, $85$ degrees, left) and worst ($20$ GeV, $85$ degrees, right) incident angle and energy combinations.
    }
    \label{fig:Sim_CoGZ}
\end{figure*}

\begin{figure*}[tbh]
    \centering
    \includegraphics[width=0.40\textwidth]{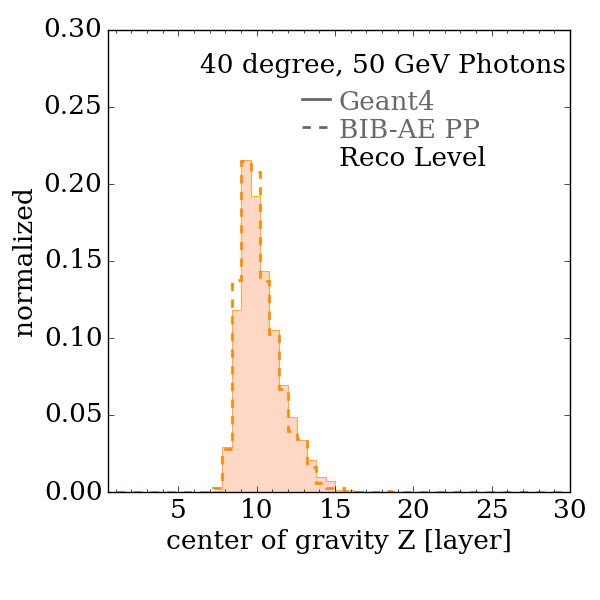}
    \includegraphics[width=0.40\textwidth]{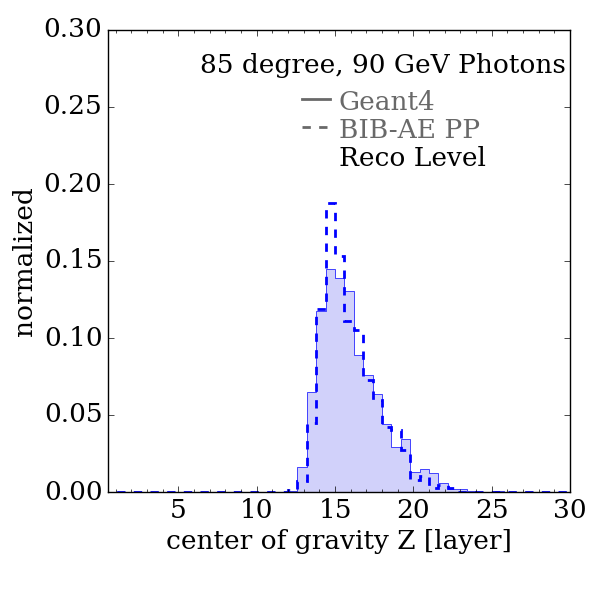}
    \caption{Reconstructed center of gravity distributions for the best ($50$ GeV, $40$ degrees, left) and worst ($90$ GeV, $85$ degrees, right) incident angle and energy combinations.
    }
    \label{fig:Reco_CoGZ}
\end{figure*}

Shower shape observables are also crucial for a generative model to capture, as they can significantly impact downstream reconstruction. We begin by reviewing the performance of the model along the depth of the calorimeter i.e along the z axis. Firstly, the center of gravity along the depth of the calorimeter, given by the first moment along the z axis, was computed and the JSD values between BIB-AE and \textsc{Geant4} distributions for each angle and energy combination are shown in Table~\ref{table:Sim_CoGZ} at both simulation and reconstruction level. At simulation level, the best ($50$ GeV, $85$ degree) and worst ($20$ GeV, $85$ degree) performing combinations of energy and angle are highlighted in bold, and the distributions shown in Figure~\ref{fig:Sim_CoGZ}. Overall, a good description of these distributions is achieved, with the most noticeable discrepancies arising from the BIB-AE tending to produce a somewhat narrower distribution than the \textsc{Geant4} counterpart. 

After reconstruction, the best and worst performing combinations of incident angle and energy are at $50$ GeV, $40$ degrees and $90$ GeV, $85$ degrees respectively. These distributions are shown in Figure~\ref{fig:Reco_CoGZ} --- in the best case (right) the BIB-AE distribution matches the \textsc{Geant4} distribution very closely, and in the worst (left) the BIB-AE again produces a slightly narrower distribution.

\subsubsection{Longitudinal Profile}\label{sec:Long}
\begin{figure*}[tbh]
    \centering
    \includegraphics[width=0.4\textwidth]{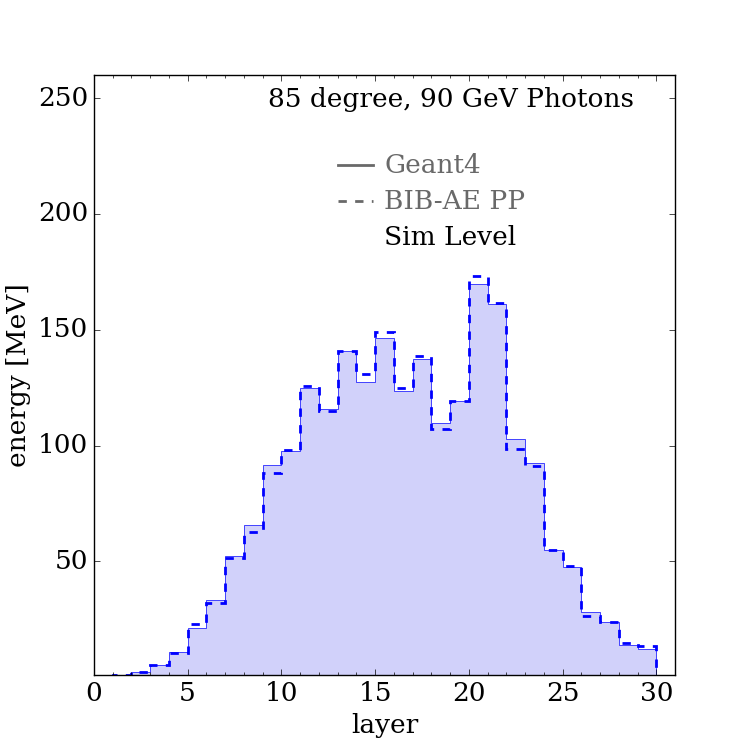} %Plots_colours/Spinal_E_40degree_20GeV.png}
    \includegraphics[width=0.4\textwidth]{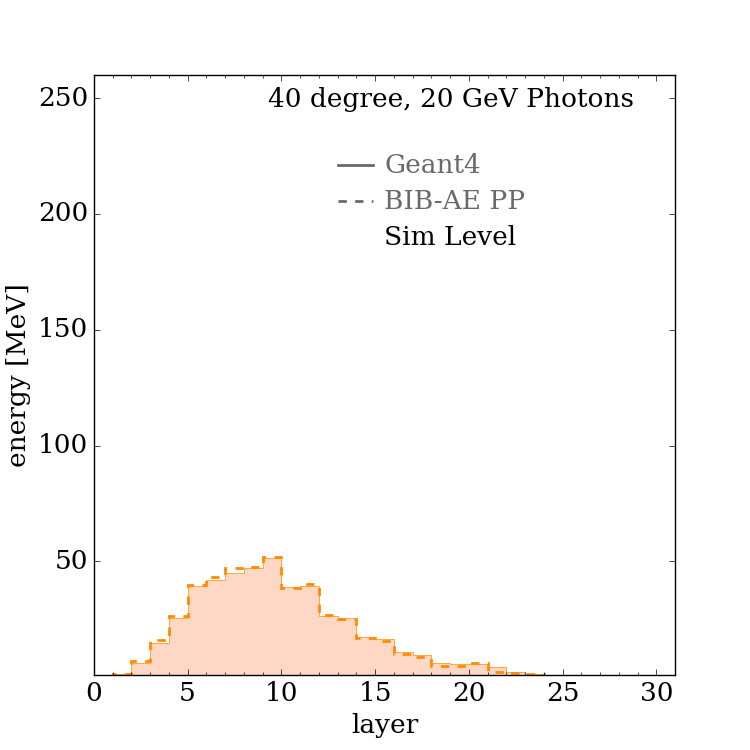}%Plots_colours/Spinal_E_40degree_50GeV.png}
    \caption{Simulation level longitudinal profiles for the best ($90$ GeV, $85$ degrees, left) and worst ($20$ GeV, $40$ degrees, right) incident angle and energy combinations.
    }
    \label{fig:Sim_Longitudinal}
\end{figure*}

\hfill\\
The second physical distribution relating to the shower development along the depth of the calorimeter that was investigated is the longitudinal profile~\footnote{Throughout this paper we refer to the longitudnial profile as the evolution of the shower into the depth of the calorimeter, rather than along the shower axis.}. The JSD values between BIB-AE and \textsc{Geant4} distributions for each combination of incident particle energy and angle are shown in Table~\ref{table:Sim_Long}. At simulation level, the best ($90$ GeV, $85$ degrees) and worst ($20$ GeV, $40$ degrees) performing combinations for energy and angle are highlighted in bold, and the distributions shown in Figure~\ref{fig:Sim_Longitudinal}. Note that the apparent discontinuity appearing around layer $20$ arises from the calibration factor applied to account for the two sampling fractions. This feature therefore becomes more pronounced for particles closer to perpendicular incidence and with higher energy, where more energy is deposited in later layers of the calorimeter. The BIB-AE performs excellently at reproducing this distribution across the combinations of energy and angle studied. This includes an impressive reproduction of the step-like features that are present in the central layers for many energy and angle combinations (see for example the $90$ GeV, $85$ degree showers in the left panel of Figure~\ref{fig:Sim_Longitudinal}), that results from the alternating layer structure of the ILD ECAL.

\begin{figure*}[tbh]
    \centering
    \includegraphics[width=0.40\textwidth]{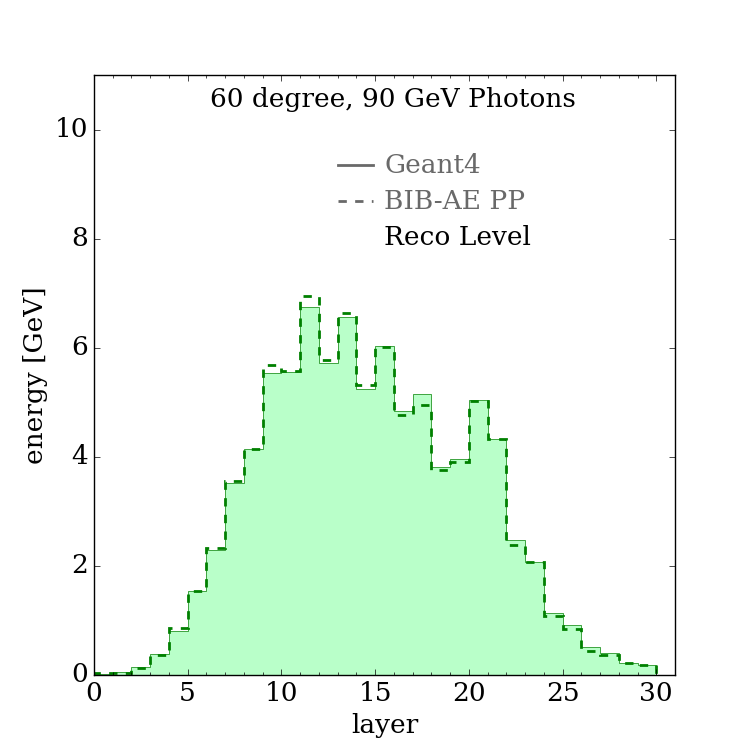}
    \includegraphics[width=0.40\textwidth]{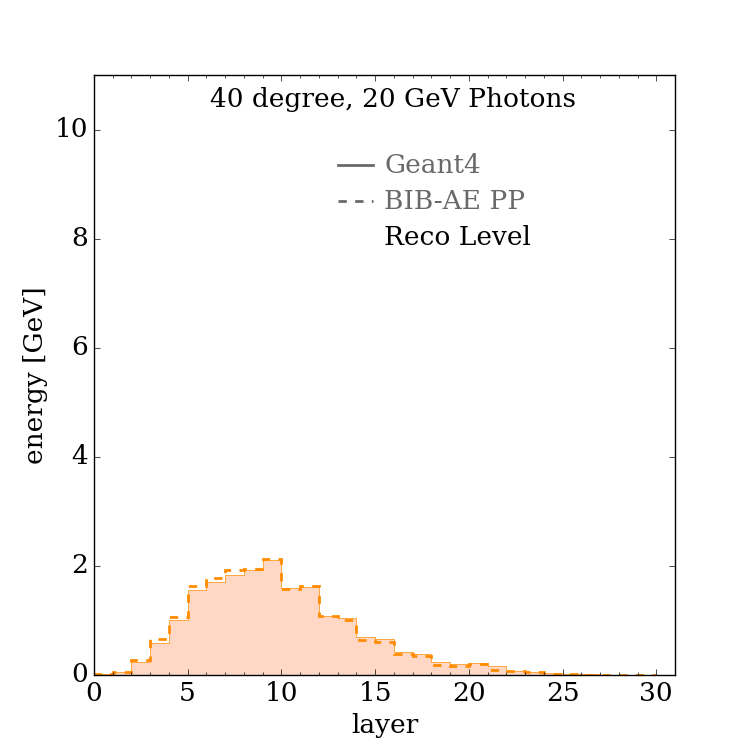}
    \caption{Reconstructed longitudinal profiles for the best ($90$ GeV, $60$ degrees, left) and worst ($20$ GeV, $40$ degrees, right) incident angle and energy combinations.
    }
    \label{fig:Reco_Longitudinal}
\end{figure*}

After reconstruction, the worst performing incident energy and angle combination is still $20$ GeV, $40$ degrees, but the best performing combination is now at $90$ GeV, $60$ degrees. The distributions for these combinations are shown in Figure~\ref{fig:Reco_Longitudinal}. The excellent level of agreement between the BIB-AE and \textsc{Geant4} distributions that was observed at simulation level is still present after the reconstruction procedure, across the range of energies and angles studied.

\begin{table*}[h!]
%\sisetup{
%separate-uncertainty=true,
%table-format=4.3(5)
%}
\setlength{\tabcolsep}{12pt}
\centering

\caption{Table showing the Jensen Shannon Distance (JSD) between the Geant4 and BIB-AE results for the simulation and reconstruction level longitudinal profiles for each combination of fixed energy and angle.}\label{table:Sim_Long}
\vspace{15pt}
\begin{center}
\let\mc\multicolumn
    \centering
        %\begin{tabular}{ll|SSS|SSS}%{@{}p{0.12\textwidth}*{4}{L{\dimexpr0.22\textwidth-2\tabcolsep\relax}}@{}}
        \begin{tabular}{ll|P{1.0cm}P{1.0cm}P{1.0cm}|P{1.0cm}P{1.0cm}P{1.0cm}}
            \toprule

                                 &&& \mc4c{Energy} \\
                                 &&& \mc4c{(GeV)}\\\midrule
                                 &&& SIM &&& REC \\
                          && 20  & 50  & 90                 & 20 & 50 & 90\\\bottomrule
Angle           & 40  &  \bf{0.025} & 0.021 & 0.023        & \bf{0.027} & 0.017 & 0.015\\
(deg.)          & 60  &  0.021 & 0.015 & 0.014             & 0.020 & 0.014 & \bf{0.009}\\
                & 85  &  0.023 & 0.014 & \bf{0.009}        & 0.023 & 0.015 & 0.011\\
                \bottomrule

\end{tabular}
\end{center}
\vspace{15pt}
\end{table*}

%\begin{table*}[h!]
%\sisetup{
%separate-uncertainty=true,
%table-format=4.3(5)
%}
%\setlength{\tabcolsep}{12pt}
%\centering

%\caption{Table showing the Jensen Shannon Distance (JSD) between the Geant4 and BIB-AE results for the sim level longitudinal profile for each combination of fixed energy and angle.}\label{table:Sim_Long}
%\vspace{15pt}
%\begin{center}
%\begin{tabular}{ll|SSS}%{@{}p{0.12\textwidth}*{4}{L{\dimexpr0.22\textwidth-2\tabcolsep\relax}}@{}}
%\toprule

%                                 &&& Energy \\
%                                 &&& (GeV)\\\midrule
%                          && 20  & 50  & 90\\\bottomrule
%Angle           & 40  &  \bf{0.025} & 0.021 & 0.023 \\
%(deg.)          & 60  &  0.021 & 0.015 & 0.014\\
%                & 85  &  0.023 & 0.014 & \bf{0.009}\\
%                \bottomrule

%\end{tabular}
%\end{center}
%\vspace{15pt}
%\end{table*}

%\begin{figure*}[tbh]
%    \centering
%    \includegraphics[width=0.4\textwidth]{Plots_Paper/Sim_plots/SpinalE/Sim_SpinalE_E90_A85.png} %Plots_colours/Spinal_E_40degree_20GeV.png}
%    \includegraphics[width=0.4\textwidth]{Plots_Paper/Sim_plots/SpinalE/Sim_SpinalE_E20_A40.png}%Plots_colours/Spinal_E_40degree_50GeV.png}
%    \caption{Sim level Longitudinal Profile
%    }
%    \label{fig:Sim_Longitudinal}
%\end{figure*}

\subsubsection{Radial Profile}\label{sec:Rad}
\begin{figure*}[tbh]
    \centering
    \includegraphics[width=0.4\textwidth]{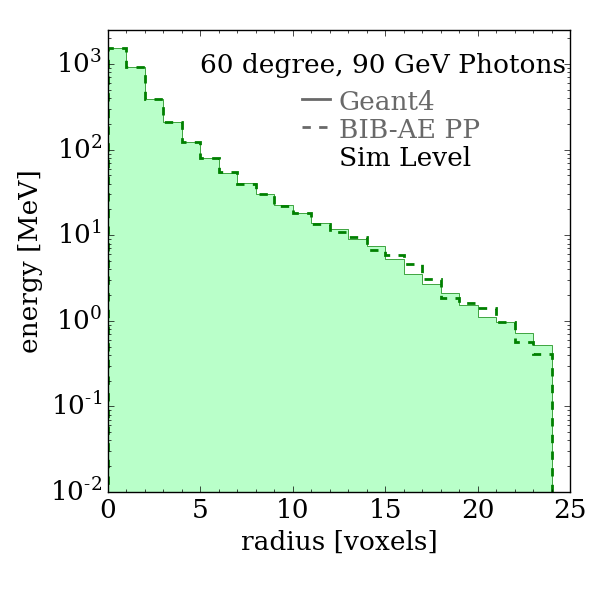}%Plots_colours/Radial_E_40degree_20GeV.png}
    \includegraphics[width=0.4\textwidth]{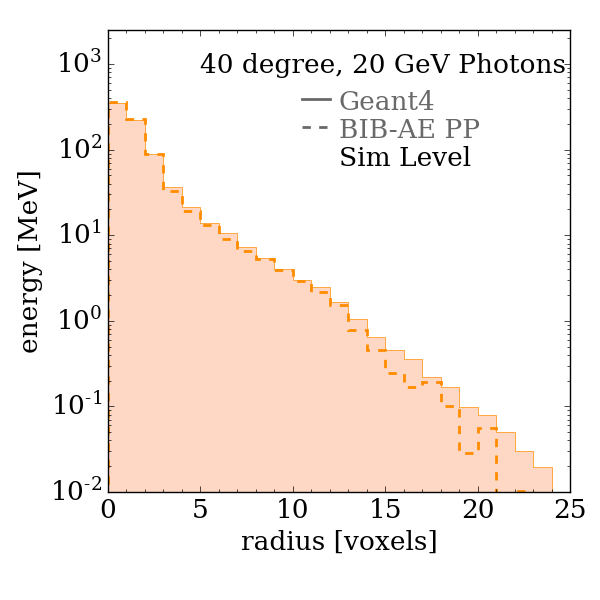}%Plots_colours/Radial_E_40degree_50GeV.png}
    \caption{Simulation level radial profiles for the best ($90$ GeV, $60$ degrees, left) and worst ($20$ GeV, $40$ degrees, right) incident angle and energy combinations. %\PJM{Replace axis label pixels with voxels}
    }
    \label{fig:Sim_Radial}
\end{figure*}

\hfill\\
The ability of the BIB-AE to capture the transversal development of showers is investigated by studying the radial profile of the showers around the principle axis. The JSD values between the BIB-AE and \textsc{Geant4} for each fixed combination of incident particle energy and angle are shown in table~\ref{table:Sim_Rad}. At simulation level, the best ($90$ GeV, $60$ degrees) and worst ($20$ GeV, $40$ degrees) performing combinations are highlighted in bold, and shown in Figure~\ref{fig:Sim_Radial}. While the BIB-AE tends to be able to capture the profile around the high energy core very well, at larger radii the performance varies. In some cases the model reproduces the distribution at larger radii with a high degree of fidelity, while in others, typically where the showers have a high degree of inclination (i.e. $40$ degree showers), the BIB-AE distributions fall off too quickly compared to the \textsc{Geant4} showers.

After reconstruction, the best and worst performing combinations of incident particle angle and energy are still $90$ GeV, $60$ degrees and $20$ GeV, $40$ degrees respectively. As at simulation level, the profile around the core of the shower is well reproduced across the board. In the best performing case (i.e $90$ GeV, $60$ degrees), even at larger radii the BIB-AE maintains a strong agreement with \textsc{Geant4}. Interestingly in the worst performing case at $20$ GeV, $40$ degrees, although a fall off at large radii is still observed, it appears to be somewhat less drastic than at simulation level. This could be significant in a more general case for distinguishing overlapping photon showers. We leave a comparison relative to \textsc{Geant4} for such a scenario to future work.

%\begin{table*}[h!]
%\sisetup{
%separate-uncertainty=true,
%table-format=4.3(5)
%}
%\setlength{\tabcolsep}{12pt}
%\centering

%\caption{Table showing the Jensen Shannon Distance (JSD) between the Geant4 and BIB-AE results for the sim level radial profile for each combination of fixed energy and angle.}\label{table:Sim_Rad}
%\vspace{15pt}
%\begin{center}
%\begin{tabular}{ll|SSS}%{@{}p{0.12\textwidth}*{4}{L{\dimexpr0.22\textwidth-2\tabcolsep\relax}}@{}}
%\toprule

 %                                &&& Energy \\
 %                                &&& (GeV)\\\midrule
 %                         && 20  & 50  & 90\\\bottomrule
 %Angle              & 40  &  \bf{0.019} & 0.014 & 0.016 \\
 %(deg.)             & 60  &  0.011 & 0.006 & \bf{0.006}\\
 %                   & 85  &  0.011 & 0.007 & 0.007\\
 %                   \bottomrule

%\end{tabular}
%\end{center}
%\vspace{15pt}
%\end{table*}

\begin{table*}[h!]
%\sisetup{
%separate-uncertainty=true,
%table-format=4.3(5)
%}
\setlength{\tabcolsep}{12pt}
\centering

\caption{Table showing the Jensen Shannon Distance (JSD) between the Geant4 and BIB-AE results for the simulation and reconstruction level radial profiles for each combination of fixed energy and angle.}\label{table:Sim_Rad}
\vspace{15pt}
\begin{center}
\let\mc\multicolumn
    \centering
        %\begin{tabular}{ll|SSS|SSS}
        \begin{tabular}{ll|P{1.0cm}P{1.0cm}P{1.0cm}|P{1.0cm}P{1.0cm}P{1.0cm}}%{@{}p{0.12\textwidth}*{4}{L{\dimexpr0.22\textwidth-2\tabcolsep\relax}}@{}}
            \toprule

                                &&& \mc4c{Energy} \\
                                &&& \mc4c{(GeV)}\\\midrule
                                &&& SIM &&& REC \\
                          && 20  & 50  & 90                     & 20 & 50 & 90\\\bottomrule
 Angle              & 40  &  \bf{0.019} & 0.014 & 0.016         & \bf{0.018} & 0.013 & 0.013\\
 (deg.)             & 60  &  0.011 & 0.006 & \bf{0.006}         & 0.010 & 0.007 & \bf{0.004}\\
                    & 85  &  0.011 & 0.007 & 0.007              & 0.010 & 0.007 & 0.008\\
                    \bottomrule

        \end{tabular}
\end{center}
\vspace{15pt}
\end{table*}

\begin{figure*}[tbh]
    \centering
    \includegraphics[width=0.40\textwidth]{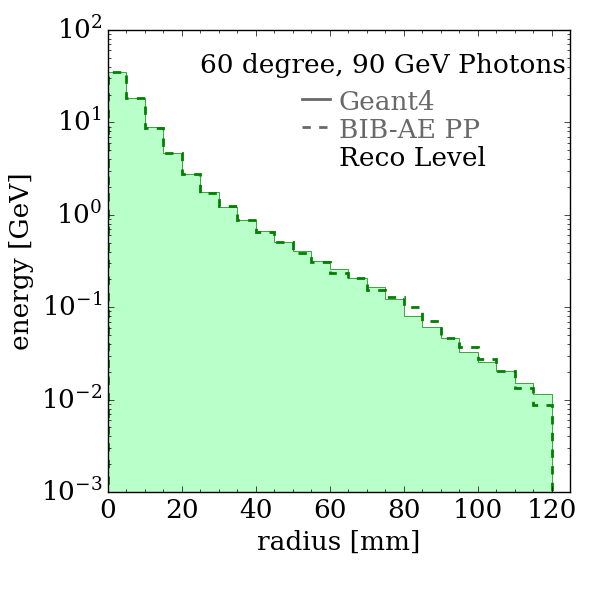}
    \includegraphics[width=0.40\textwidth]{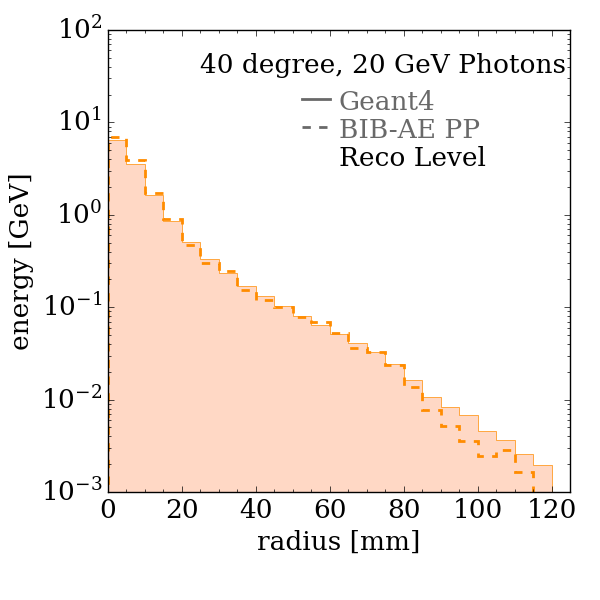}
    \caption{Reconstructed radial profiles for the best ($90$ GeV, $60$ degrees, left) and worst ($20$ GeV and $40$ degrees, right) incident angle and energy combinations.
    }
    \label{fig:Reco_Radial}
\end{figure*}

%\subsection{Reconstruction level results}\label{sec:RecoResults}

\subsection{Computational Performance}\label{sec:COmpPerf}

The suitability of a generative model as a surrogate simulator ultimately comes down to its inference time per shower. To this end, the generation time per sample is benchmarked on both CPU and GPU hardware. Table~\ref{table:bench} shows the average time to generate a shower with energy and angle uniformly distributed in the 10-100 GeV range and 90-30 degree range respectively using \textsc{Geant4} and the BIB-AE. The generative model offers a significant speedup relative to \textsc{Geant4}, reaching up to three orders of magnitude on a GPU.

\begin{table*}[h!]
\sisetup{
separate-uncertainty=true,
table-format=4.3(5)
}
\centering

\caption{Comparison of the computational performance of the BIB-AE generator and \textsc{Geant4} on a single core of an Intel\textsuperscript{\tiny\textregistered} Xeon\textsuperscript{\tiny\textregistered} CPU E5-2640 v4 (CPU) and an NVIDIA\textsuperscript{\tiny\textregistered} A100 with 40~GB of memory (GPU). For the BIB-AE, the best performing batch size is selected. The value shown is the mean obtained for a set of 2000 showers with uniform energy from $10-100$ GeV and $30-90$ degrees, with error arising from the standard deviation.}\label{table:bench}
\vspace{15pt}
%\begin{tabular}{ll|Sr}
%\begin{tabular}{P{2.5cm}P{2.5cm}|P{2.5cm}P{2.5cm}}
\begin{tabular}{ll|P{2.5cm}r}
\toprule
Hardware & Simulator &  {Time / Shower [ms]} & Speed-up \\\midrule
CPU & \textsc{Geant4} & 4417 $\pm$ 83 & $\times 1$ \\
    & & & \\
    & BIB-AE & 362 $\pm$ 2 & $\times 12$ \\
    & & &\\
GPU & BIB-AE & 4.32 $\pm$ 0.09 & $\times 1022$ \\\bottomrule
\end{tabular}
\vspace{15pt}
\end{table*}

\section{Conclusion}\label{sec:Conclusion}
Generative models show potential to provide powerful tools for fast simulation, and to significantly reduce the computational resources required by experiments in particle physics. The contribution of this paper is two fold. In the first instance, we generalise the BIB-AE architecture, one of the most powerful generative models for calorimeter simulation, to a multi-parameter conditioning scenario in a highly granular calorimeter. A key challenge for extending simulation tools based on generative models into such a multi-parameter conditioning scenario arises from the requirement to cover an increased phase space. This means a generative model not only needs a robust conditioning scheme, but also requires sufficient capacity and capability. Scaling of models designed for learning on regular grids must also be given some forethought, as larger grid sizes will be required for showers with varying angle of incidence. In the second instance we perform a detailed study into the effects of particle flow reconstruction on the performance of a generative model for the first time in such a highly granular detector. We demonstrate the possibility to design a simulator that retains a strong performance after reconstruction --- it will be the physics performance after this processing that will provide the ultimate means of judging the suitability of a surrogate simulator.

The physics performance of the BIB-AE model was shown to be strong across a range of physics observables. From a conditioning perspective, the energy response was shown to be particularly strong- thanks in part to the per-shower re-scaling afforded by the use of a normalising flow to learn the total energy deposited in the active regions of the calorimeter. This approach helps to make the energy conditioning performance of the model more reliable than was the case in previous versions~\cite{GettingHigh}. The angular response was somewhat weaker in comparison, but still provides a good description overall. 

%\PJM{ Do we want to comment on other work looking at angular conditioning here? } \gk{yes}
%In relation to other work, little focus has been given to conditioning generative models for highly granular calorimeter shower simulation on the angle of incidence, with the notable exception of~\cite{Khattak:2021ndw}. However, in that case no direct study of the angular conditioning performance of the model was made.
The BIB-AE retained the ability to learn the cell energy spectrum well across a range of angles and energies, as well as reproducing shower shapes with a high degree of fidelity. The most noticeable differences in the distributions appear in the radial profile for more inclined showers and at larger radii.

The improved computational potential of generative models relative to \textsc{Geant4}, which is by this point well established in the literature, remains true for the BIB-AE architecture in this extended setup. A speed-up of up to three orders of magnitude relative to \textsc{Geant4}, aligns with previous results for this model~\cite{GettingHigh}~\cite{HBFS}.

For future work, a crucial step will be to develop general methods to handle irregularities in the detector geometry. This will be necessary to allow the incident position on a detector surface to be varied, and entire detector subsystems to be handled. This step would also allow the effects of reconstruction to be studied in more general environments where overlapping showers are present. Studying the physics performance of generative models after reconstruction is an essential step, which will ultimately be necessary to evaluate the suitability of any generative model for calorimeter simulation.

%\section*{References}

%\newpage

\ack
This research was supported in part through the Maxwell computational resources operated at Deutsches Elektronen-Synchrotron DESY, Hamburg, Germany. This project has received funding from the European Union’s Horizon 2020 Research and Innovation programme under Grant Agreement No 101004761. 
SD is funded by the Deutsche Forschungsgemeinschaft under Germany’s Excellence Strategy – EXC 2121  Quantum Universe – 390833306. 
We acknowledge support via the KISS consortium (05D23GU4, 13D22CH5) funded by the German Federal Ministry of Education and Research BMBF in the ErUM-Data action plan.

\section*{References}

\bibliographystyle{iop-num-mod}
\bibliography{main.bib}

\end{document}